\documentclass[prb,superscriptaddress,twocolumn,showpacs,amsmath,amssymb,floatfix]{revtex4}

\usepackage{graphicx}
\newcommand{\PEC}{(Pr$_{1-y}$Eu$_y$)$_{0.7}$Ca$_{0.3}$CoO$_3$}
\newcommand{\osix}{$^{16}$O}
\newcommand{\oeight}{$^{18}$O}
\newcommand{\fref}[1]{Fig.~\ref{#1}}
\newcommand{\coiii}{Co$^{3+}$}
\newcommand{\coiiii}{Co$^{4+}$}
\newcommand{\eg}{$e_g$}
\newcommand{\tg}{$t_{2g}$}
\newcommand{\tgeg}[2]{$t_{2g}^#1e_g^#2$}
\newcommand{\JKm}{~J/(K\thinspace mol)}
\newcommand{\muB}{~$\mu_B$}
\newcommand{\muBco}{~$\mu_B/$Co}
\newcommand{\oC}{$^\circ$C}
\newcommand{\mREonH}[1]{$M_\mathrm{#1}(H)$}

\begin{document}

\title{Phase diagram and isotope effect in (Pr$_{1-y}$Eu$_{y}$)$_{0.7}$Ca$_{0.3}$CoO$_3$
cobaltites exhibiting spin-state transitions}

\author{A. V. Kalinov}\email[E-mail: ]{kalinov@vei.ru}
\affiliation{All-Russian Electrical Engineering Institute,
Krasnokazarmennaya Str.\ 12, 111250 Moscow, Russia}

\affiliation{{II.}\ Physikalisches Institut, Universit\"at zu
K\"oln, Z\"ulpicher Str.\ 77, 50937 K\"oln, Germany}

\author{\framebox{O. Yu. Gorbenko}}
\affiliation{Department of Chemistry, Moscow State University,
119991 Moscow, Russia}

\author{A. N. Taldenkov}
\affiliation{Institute of Molecular Physics, Russian Research
Center ``Kurchatov Institute'', Kurchatov Square 1, 123182
Moscow, Russia}

\author{J. Rohrkamp}
\affiliation{{II.}\ Physikalisches Institut, Universit\"at zu
K\"oln, Z\"ulpicher Str.\ 77, 50937 K\"oln, Germany}

\author{O.~Heyer}
\affiliation{{II.}\ Physikalisches Institut, Universit\"at zu
K\"oln, Z\"ulpicher Str.\ 77, 50937 K\"oln, Germany}

\author{S. Jodlauk}
\affiliation{{II.}\ Physikalisches Institut, Universit\"at zu
K\"oln, Z\"ulpicher Str.\ 77, 50937 K\"oln, Germany}

\affiliation{Institut f\"{u}r Kristallographie, Universit\"{a}t
zu K\"{o}ln, Z\"{u}lpicher Str.\ 49b, 50674 K\"{o}ln, Germany}

\author{N. A. Babushkina}
\affiliation{Institute of Molecular Physics, Russian Research
Center ``Kurchatov Institute'', Kurchatov Square 1, 123182
Moscow, Russia}

\author{L. M. Fisher}
\affiliation{All-Russian Electrical Engineering Institute,
Krasnokazarmennaya Str.\ 12, 111250 Moscow, Russia}

\author{A. R. Kaul}
\affiliation{Department of Chemistry, Moscow State University,
119991 Moscow, Russia}

\author{A.~A.~Kamenev}
\affiliation{Department of Chemistry, Moscow State University,
119991 Moscow, Russia}

\author{T. G. Kuzmova}
\affiliation{Department of Chemistry, Moscow State University,
119991 Moscow, Russia}

\author{D. I. Khomskii}
\affiliation{{II.}\ Physikalisches Institut, Universit\"at zu
K\"oln, Z\"ulpicher Str.\ 77, 50937 K\"oln, Germany}

\author{K. I. Kugel}
\affiliation{Institute for Theoretical and Applied
Electrodynamics, Russian Academy of Sciences, Izhorskaya Str.\
13, 125412 Moscow, Russia}

\author{T. Lorenz}
\affiliation{{II.}\ Physikalisches Institut, Universit\"at zu
K\"oln, Z\"ulpicher Str.\ 77, 50937 K\"oln, Germany}

\begin{abstract}
We present the study of magnetization, thermal expansion,
specific heat, resistivity, and a.c. susceptibility of
(Pr$_{1-y}$Eu$_y$)$_{0.7}$Ca$_{0.3}$CoO$_3$ cobaltites. The
measurements were performed on ceramic samples with $y = 0.12 -
0.26$ and $y = 1$. Based on these results, we construct the
phase diagram, including magnetic and spin-state transitions.
The transition from the low- to intermediate-spin state is
observed for the samples with $y > 0.18$, whereas for a lower
Eu-doping level, there are no spin-state transitions, but a
crossover between the ferromagnetic and paramagnetic states
occurs. The effect of oxygen isotope substitution along with Eu
doping on the magnetic/spin state is discussed. The
oxygen-isotope substitution ($^{16}$O by $^{18}$O) is found to
shift both the magnetic and spin-state phase boundaries to
lower Eu concentrations. The isotope effect on the spin-state
transition temperature ($y > 0.18$) is rather strong, but it is
much weaker for the transition to a ferromagnetic state ($y <
0.18$). The ferromagnetic ordering in the low-Eu doped samples
is shown to be promoted by the Co$^{4+}$ ions, which favor the
formation of the intermediate-spin state of neighboring
Co$^{3+}$ ions.
\end{abstract}

\date{\today}

\maketitle

\section{Introduction}
Perovskite-based mixed-valence oxides with the general formula
$R_{1-x}A_xTM$O$_3$, where $R$, $A$, and $TM$ are rare-earth,
alkali-earth, and transition-metal elements, respectively, have
attracted a lot of interest due to a rich variety of their
electronic and magnetic states. These transition-metal oxides
exhibit different ordering phenomena and phase transitions,
e.g., antiferromagnetic (AFM) and/or ferromagnetic (FM) order,
charge and/or orbital orderings, metal-insulator
transitions.\cite{imada.rmp98} The interplay of different
degrees of freedom and different types of ordering is a very
important ingredient in determining the properties of such
strongly correlated electron systems. These effects become
especially interesting in doped materials, which may exhibit a
tendency to phase separation and to the formation of
inhomogeneous states.\cite{khomskii.pb00, kagan.ufn01,
dagotto.03}

In addition to these phenomena common to most doped $TM$
oxides, cobaltites have an extra ``degree of freedom'', namely,
the \coiii\ ions in them may occur in different spin states
(belonging to different multiplets): low-spin (LS; spin $S =
0$; \tgeg{6}{0}), intermediate-spin (IS; $S = 1$; \tgeg{5}{1}),
or high-spin (HS; $S = 2$; \tgeg{4}{2}) states with the
possibility of spin-state transitions (SST) between them
caused, e.g., by temperature, pressure, or
doping.\cite{jonker.vansanten.53, goodenough.jap65,
senarisrodriguez.jssc95, asai.jpsj98, saitoh.prb97,
tokura.prb98} The existence of a spin-state change indicates
that the difference of the electronic energies, $\Delta E$,
between these states is rather small. The most prominent
example is LaCoO$_3$, which was actively studied and
controversially debated for more than 50
years.\cite{jonker.vansanten.53, goodenough.jap65,
senarisrodriguez.jssc95, korotin.prb96, asai.jpsj98,
saitoh.prb97, tokura.prb98, yamaguchi.prb97, kobayashi.prb00,
sato.jpsj08, zobel.prb02, berggold.prb08} It is generally
agreed that \coiii\ ions in LaCoO$_3$ are in the LS state at
low temperatures. Above approximately 25 K, a higher-spin
state, either IS or HS, becomes thermally populated affecting
various physical properties, e.g., magnetic susceptibility
$\chi$ or thermal expansion $\alpha$, which both exhibit
pronounced peaks in their temperature
dependence.\cite{zobel.prb02, baier.prb05, berggold.prb08} The
susceptibility is naturally affected because the excited spin
state (IS or HS) induces a significant increase in the
magnetization. The thermal expansion is affected due to the
different ionic radii: the LS \coiii\ ions with empty \eg\
orbitals are significantly smaller than the IS or HS \coiii\
ions with partially filled \eg\ orbitals. It should be noticed
that there are some structural (neutron diffraction)
data,\cite{tong.jpsj09, tsubouchi.prb02, fujita.jpsj04} which
show almost no change in the average Co--O bond lengths from 4
to 300~K for A-cite substituted cobaltites with the SST.
Instead, a distortion of CoO$_6$ octahedra takes place upon the
SST giving rise to an almost unchanged unit cell
volume.\cite{tong.jpsj09} In Ref.~\onlinecite{tsubouchi.prb02}
the nearly constant average Co--O bond-length on the SST is
attributed to the stronger covalency in the high-temperature
phase.

In undoped LaCoO$_3$, there is not yet a general consensus
whether the excited state is IS or HS. For example, our earlier
thermal-expansion and magnetic measurements \cite{zobel.prb02}
were interpreted in terms of a temperature-induced excitation
of a triplet state, which may suggest that this excited state
is an IS \coiii\ ($S = 1$). However, in the presence of a
strong spin-orbit coupling, typical for Co with partially
filled \tg\ levels, it is the HS \coiii\, which has a triplet
ground state with a total effective moment $J =
1$.\cite{podlesnyak.prl06, haverkort.prl06, ropka.prb03,
haverkort.phd} This led to the conclusion that the thermally
excited state in undoped LaCoO$_3$ is most probably the HS
state \cite{podlesnyak.prl06, noguchi.prb02, haverkort.prl06}
in contrast to many earlier claims.\cite{korotin.prb96,
tokura.prb98, zobel.prb02, baier.prb05} This conclusion is
still disputed.\cite{phelan.prb08,phelan.prb09}

In hole-doped cobaltites, the situation is even more complex.
One can argue that due to the presence of low-spin \coiiii, it
would be indeed the IS state of \coiii, which is created close
to a hole;\cite{podlesnyak.prl06, podlesnyak.prl08,
phelan.prb08} this would allow for a free motion of a hole
within a cluster of LS \coiiii\ and neighboring IS \coiii\
ions.\cite{berggold.prb08, maignan.prl04,kriener.prb04} In the
itinerant case, the stabilization of a $J = 1$ state due to
spin-orbit interaction would also become questionable. Thus,
based on these arguments, as well as on the results of
Ref.~\onlinecite{podlesnyak.prl08}, we conclude that the most
plausible scenario is that the doped holes promote a certain
amount of \coiii\ to an IS state, with holes moving in a
respective \coiiii/\coiii\ cluster. Due to the usual double
exchange \cite{zener.pr51, degennes.pr60} such clusters would
then be ferromagnetic.

The SST in $R$CoO$_3$ is strongly affected by both heterovalent
and isovalent doping at the $R$ site. The heterovalent doping
($R_{1-x}$$A_x$CoO$_3$; $A$ = Ba, Sr, and Ca) causes hole
doping and chemical pressure (arising from the change of the
ionic radii in the $R_{1-x}A_x$ complex) and stabilizes a
magnetic (IS or HS) state of
\coiii.\cite{kriener.prb04,kriener.prb09} For the relatively
large atomic species $R$ = La and $A$ = Ba or Sr, the
nonmagnetic (LS) insulating ground state of $R$CoO$_3$ changes
to a ferromagnetic metal at high enough hole concentration,
which is due to the enhancement of the double exchange
interaction caused by the decrease of $\Delta E$ and an
increase in the number of electrons at the \eg\ orbitals. The
isovalent doping with a smaller rare-earth, i.e., chemical
pressure without changing the Co valence, is usually realized
by introducing other trivalent rare-earth ions $R^{3+}$. In
that case, the LS state of \coiii\ ions is stabilized and the
spin-state transition is shifted to a higher temperature. For
example, the energy gap between the LS and the excited spin
state increases from about $\Delta E = 185$~K for $R$ = La to
$\Delta E \ge 2000$~K for $R$ = Eu.\cite{baier.prb05} In
addition, with decreasing ionic radius in the lanthanide
series, the structure changes from rhombohedral in LaCoO$_3$ to
orthorhombic for $R$ = Pr, Nd, and Eu.\cite{berggold.prb08} The
changes in the ionic radius can be also fine-tuned by mixing
$R$ elements with different ionic radii. In this situation, the
partial substitution of $R^{3+}$ by a smaller $R'^{3+}$ in
($R_{1-y}R'_y$)$_{1-x}A_x$CoO$_3$ systematically increases the
chemical pressure, which should enhance the crystal field
splitting and therefore stabilize the low-spin
state.\cite{baier.prb05, fujita.jpsj05} Thus, small changes in
the lattice characteristics may critically determine the
physical properties of cobaltites.

Phenomena, for which the crystal lattice plays a significant
role, usually show a strong isotope effect, and vice versa
studies of the isotope effect can yield important information
about the underlying mechanisms. One famous example is the
isotope effect in conventional superconductors, which led to
the conclusion that in these systems the electron pairing is
caused by the electron-phonon interaction. In some cases,
especially if the system is close to a crossover between
different states, an isotope substitution can drastically
change the properties of the system. For example, in
manganites, one can even induce a metal-insulator transition by
substituting \osix\ by \oeight.\cite{babushkina.nat98}
Moreover, the isotope substitution can be also used for
fine-tuning the behavior of a system, without stronger
disturbances such as those caused, e.g., by doping (which
introduce extra disorder). In this sense, the isotope
substitution is an even ``softer'' way to control the behavior
of a material. For cobaltites with the SST, we can expect a
particularly pronounced isotope effect because of the strong
involvement of the lattice in it due to the fact that the ionic
radii of different spin states are very different: the ionic
radius of HS \coiii\ ($S=2$) is by about 15\% larger than that
for the LS state ($S=0$).

Indeed, the first measurements \cite{wang.prb06, wang.prb06.2}
have already demonstrated the existence of an oxygen isotope
effect in cobaltites with spin-state transitions. With
increasing $x$ in (Pr$_{1-x}$Sm$_x$)$_{0.7}$Ca$_{0.3}$CoO$_3$,
a crossover from a ferromagnetic metal to an insulator
exhibiting a spin-state transition was found. The authors
observed \cite{wang.prb06, wang.prb06.2} that the oxygen
isotope substitution strongly shifts the spin-state transition
temperature $T_\mathrm{SS}$ in the insulating phase, but only
slightly affects the ferromagnetic transition point
$T_\mathrm{FM}$. This contrasting behavior in the two phases
was explained by the occurrence of static Jahn-Teller (JT)
distortions in the insulating phase and the absence of them in
the metallic phase. Here, a remark should be made. The \coiii\
ions in the IS state have formally the \tgeg{5}{1}
configuration, i.e., they indeed could exhibit a pronounced JT
effect. However, if the corresponding \eg\ electrons are
itinerant, which seems to be the case in our system, one should
not expect a strong JT effect. Therefore, we do not think that
the coupling to the lattice in this case has predominantly a JT
character, as assumed in Ref.~\onlinecite{wang.prb06.2};
rather, simply the difference of the ionic sizes of the LS and
the IS or HS states of \coiii\ should play the most important
role here.

In the present study, we tried to clarify the nature of the
strong difference in the character of the magnetic/spin-state
transition for the ``metallic'' and ``insulating'' ground
states. For this, in particular, we used the oxygen isotope
exchange to influence the crystal-field splitting without
additional effects of static distortions. For these purposes,
the \PEC\ ($y = 0.12 - 0.26$) series with \osix\ and \oeight\
were studied. This series was chosen to span the range of
ground states from metal-like to insulating ones. We measured
d.c.\ magnetization, low-frequency a.c.\ magnetic
susceptibility, thermal expansion, specific heat, and
resistivity. Based on these results, we have constructed a
detailed phase diagram representing different magnetic and spin
states. We find that with increasing Eu doping, that is, with
decreasing average size of (Pr$_{1-y}$Eu$_y$) or increasing
chemical pressure, the ground state of the compound changes
from a ``ferromagnetic metal'' to a ``weakly-magnetic
insulator'' at $y \approx y_\mathrm{th} = 0.18$. A pronounced
SST is present in the insulating ground state (in the samples
with $y > y_\mathrm{th}$). The metallic ground state (in the
samples with $y < y_\mathrm{th})$ has completely different
magnetic properties, without any indications of a
temperature-induced spin-state transition. There is even no
well-defined magnetic transition as a function of temperature
for this phase. Instead, a smooth crossover to a paramagnetic
state takes place. The isotope exchange affects strongly the
SST temperature in the insulating regime, but has only a
marginal effect on the crossover temperature in the samples
with the metallic ground state. Possible implications of the
spin ordering of \coiiii\ ions for the spin-state transition
are also discussed.

\section{Experimental details}

\subsection{Samples}

\begin{figure}[!tb]
\includegraphics{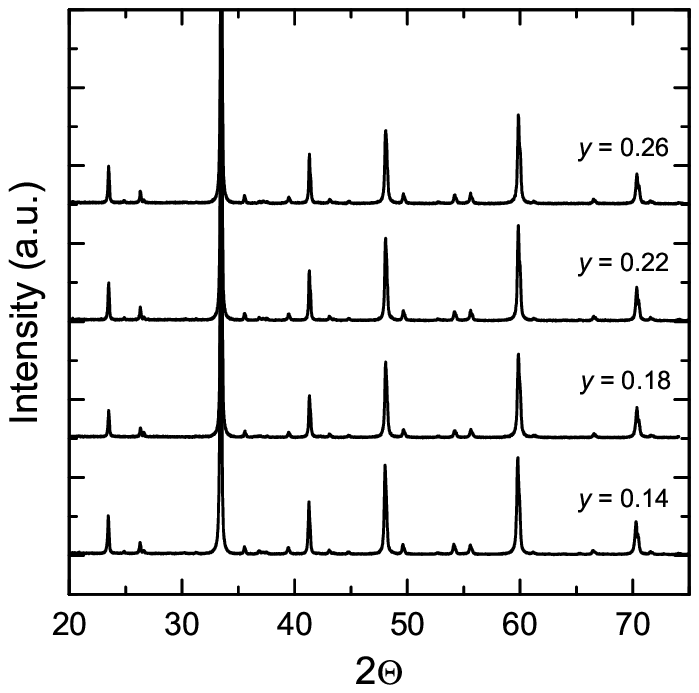}
\caption{\label{xrd} Representative X-ray diffraction data for
the \PEC\ samples with $y=0.14$, 0.18, 0.22 and 0.26.}
\end{figure}

In our measurements, we used ceramic \PEC\ samples with both
\osix\ and \oeight. A series of samples with $y = 0.12 - 0.26$
was synthesized. The substitution of Pr by Eu results in the
decrease of the average ionic radius of the rare-earth
combination. This enhances the static crystallographic
distortion and is followed by an increase of the crystal-field
splitting that should result in a stabilization of the low-spin
state against the temperature-activated spin-state transition
to the intermediate- or high-spin state. The Pr and Eu content
in \PEC\ was chosen from the data on ionic radii from Shannon's
tables \cite{shannon.76} for nine-fold coordination of
rare-earth and alkaline-earth cations, which is usually
accepted for cobaltites with orthorhombic symmetry of the
perovskite structure. The chosen stoichiometries span the mean
rare-earth radius $\langle r_A \rangle$ in the range from
1.172~\AA\ to 1.164~\AA, for $y = 0.12 - 0.26$. The end-member
of this series, Eu$_{0.7}$Ca$_{0.3}$CoO$_3$, was also studied
and some results obtained earlier \cite{berggold.prb08,
baier.prb05} on $R$CoO$_3$ with $R$ = La, Pr, Nd, Eu are
referred below for comparison. The \PEC\ samples were prepared
by the ``paper synthesis'' technique:\cite{balagurov.prb99}
ash-free paper filters were impregnated with an aqueous
solution of a mixture of nitrates of the corresponding metals,
dried at 120\oC, and then burnt. The resultant powder was
annealed at $T = 700$\oC\ for 2~h for decarbonization. After
pressing, the ceramics were sintered at $T = 1000$\oC\ for
100~h. The X-ray analysis with Rigaku SmartLab diffractometer
(using Cu$K\alpha$ irradiation, $\lambda=1.542$~\AA)
demonstrated that the samples are single-phase without
observable admixture of impurity phases (\fref{xrd}). All the
data were indexed in the orthorhombic lattice $Pnma$. No
structure modification was observed with Eu-doping. The unit
cell volume was found to decrease from 216.5~\AA$^3$ for
$y=0.12$ to 215.7~\AA$^3$ for $y=0.26$  that agrees  well  with
the data of Fujita et al. \cite{fujita.jpsj04, fujita.jpsj05}
obtained by both X-ray and neutron diffraction techniques. The
unit cell volume for our series is about 0.15\% smaller as
compared to the (Pr$_{1-y}$Sm$_y$)Ca$_{0.3}$CoO$_3$ series
which is naturally attributed to the smaller ionic radius of
europium as compared to samarium.

The nominal content of Eu along the series is shown in
Table~\ref{table1}. Hereafter, these samples are referred to as
Eu$_y$-$n$, where $y$ and $n = 16$ or 18 denote the Eu content
and the oxygen isotope, respectively. We performed measurements
for all the samples of the series, but in most cases only the
results for the samples with $y=0.14$, 0.18, 0.22, and 0.26 are
shown because for samples, which differ in $y$ by only 0.02,
the data in almost all cases overlap when the oxygen isotope is
exchanged, i.e., the data of Eu$_{0.12}$-18 overlap with those
Eu$_{0.14}$-16 and so on.

\begin{table}
\caption{\label{table1} The list of \PEC\ samples. Nominal Eu
compositions are shown along with the oxygen isotope and the
average ionic rare-earth radii calculated on the basis of
Ref.~\onlinecite{shannon.76}.}
\begin{ruledtabular}
\begin{tabular}{cccc}
\omit&Nominal&Oxygen&\omit\\
Acronym&$y$&isotope&$\langle r_A\rangle$~(\AA)\\
\hline
Eu$_{0.12}$-16/18&0.119&\osix/\oeight&1.1720\\
Eu$_{0.14}$-16/18&0.137&\osix/\oeight&1.1709\\
Eu$_{0.16}$-16/18&0.159&\osix/\oeight&1.1696\\
Eu$_{0.18}$-16/18&0.179&\osix/\oeight&1.1684\\
Eu$_{0.20}$-16/18&0.199&\osix/\oeight&1.1678\\
Eu$_{0.22}$-16/18&0.219&\osix/\oeight&1.1661\\
Eu$_{0.24}$-16/18&0.239&\osix/\oeight&1.1649\\
Eu$_{0.26}$-16/18&0.259&\osix/\oeight&1.1637\\
Eu$_{0.7}$Ca$_{0.3}$CoO$_3$&1.0&\osix&1.12\\
\end{tabular}
\end{ruledtabular}
\end{table}

The \osix\ to \oeight\ isotope substitution was carried out by
annealing the samples in oxygen at 950\oC\ for 200~h. The gas
pressure was equal to 1~bar. Two samples of rectangular shape
prepared from the same pellet were simultaneously annealed, one
in \osix$_2$ and the other in \oeight$_2$ (93\% of
\oeight$_2$). The final enrichment of the samples with \oeight\
was 92\%, as determined from the weight change. The enrichment
process is described in detail
elsewhere.\cite{babushkina.jap98} The similarity of the oxygen
isotope composition in the sample to that in the gas medium
indicated that a thermodynamic equilibrium was achieved during
annealing and, hence, the difference in the diffusion rates of
the oxygen isotopes did not significantly affect the results of
the investigation. It should also be noted that the mass of a
sample annealed in \osix$_2$ remained unchanged (within the
experimental error) during the prolonged heat treatment. Thus,
we can conclude that the annealing procedure does not change
the oxygen stoichiometry in the compounds under study. The
measurement of the real oxygen stoichiometry via iodometric
titration in oxygen-exchanged samples does not provide a
relevant accuracy because of the small amount of sample mass.
However, the data obtained for samples of
Pr$_{0.7}$Sr$_{0.3}$CoO$_3$ [\onlinecite{brinks.jssch99}] and
La$_{0.7}$Sr$_{0.3}$CoO$_3$ [\onlinecite{jonker.vansanten.53,
mineshige.jssch96}] prepared under normal oxygen pressure show
that the expected oxygen deficiency should be about $0.02 -
0.04$. The comparison of the transition temperatures for
closely related (PrSm)$_{0.7}$Ca$_{0.3}$CoO$_3$ samples
annealed under oxygen pressure of 1 bar
[\onlinecite{wang.prb06}] and of 60 bar
[\onlinecite{fujita.jpsj04, fujita.jpsj05}] suggests that the
presumed oxygen deficiency hardly affects the physical
properties.

\subsection{Techniques}

The magnetization of the samples was measured in a vibrating
sample magnetometer insert to the Quantum Design PPMS in the
magnetic field range up to 14~T at temperatures 2 -- 300~K. For
some measurements at low magnetic fields, a Quantum Design
SQUID (MPMS) was used. The magnetic susceptibility $\chi(T)$
was measured in an a.c.\ magnetic field with the frequency of
667~Hz and an amplitude of 5~Oe. The electrical resistance
$R(T)$ of the samples was measured using the four-probe
technique [a two-probe method was used for $R(T) > 1 \Omega$]
in the temperature range from 5 to 330~K in magnetic fields up
to 4~T; the magnetic field was directed parallel to the
transport current. The specific heat $C_p$ was measured using
the Quantum Design PPMS by the two-tau relaxation technique
from 2 -- 300~K in magnetic fields up to 14~T. High-resolution
measurements of the linear thermal expansion coefficient
$\alpha= (1/L) dL/dT$ were performed on heating from 4 to 180~K
using a home-built capacitance dilatometer.

\section{Experimental Results}

\begin{figure*}[!tb]
\includegraphics{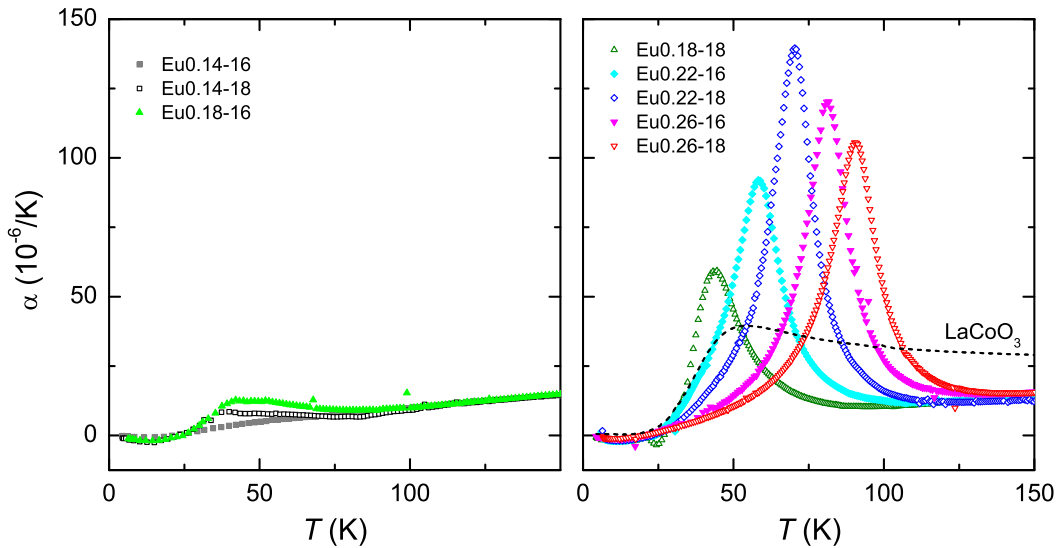}
\caption{(Color online)\label{alpha} Thermal expansion $\alpha$
of \PEC\ as a function of temperature for low-distorted (LD,
left panel) and highly-distorted (HD, right panel) samples.
Only HD samples (more Eu) demonstrate the pronounced
spin-state-transition anomaly. For comparison, the anomalous
thermal expansion arising from the spin-state transition of
LaCoO$_3$ is also shown (dashed curve).}
\end{figure*}

The measurements of thermal expansion $\alpha$ are illustrated
in \fref{alpha}. Here and further on, we denote the samples
according to table~\ref{table1}. Obviously, the data can be
split in two groups: the ``highly distorted'' (HD) samples
(larger $y$) show pronounced anomalies, which are almost
completely absent in the ``less distorted'' (LD) samples. The
borderline between both groups is located at $y_\mathrm{th}
\approx 0.18$ and depends on the oxygen isotope or, in other
words, the (almost) absence or presence of this strong anomaly
can be switched by exchanging the oxygen isotope from \osix\ to
\oeight. As will become clearer in the following, we can
attribute these large anomalies to a spin-state transition of
the \coiii\ ions, which in this series manifests itself as a
first-order phase transition. The fact that the anomalies are
comparatively broad (FWHM is about 25~K) indicates that there
is a large temperature range of phase coexistence and there are
also indications that the different phases even coexist up to
the highest and lowest measured temperatures (see below). This
is most probably also the cause, why the residues of these
anomalies are still visible in some of the LD samples (left
panel of \fref{alpha}). As mentioned above, a strong lattice
expansion is expected to occur at a SST due to the
significantly different ionic radii of the LS and the
higher-spin states of \coiii. We would like to stress, however,
that the behavior here is very different from that of LaCoO$_3$
\cite{zobel.prb02, baier.prb05, radaelli.prb02} or other
RCoO$_3$ compounds.\cite{berggold.prb08} The anomalous
expansion due to the SST of LaCoO$_3$, which for comparison is
also shown in \fref{alpha}, has the typical form of a Schottky
anomaly meaning that the SST is related to the thermal
population of an energetically higher-lying higher-spin state,
while here we are dealing with a real (but broadened)
first-order phase transition in the thermodynamic sense. This
explains the very symmetric shape of the $\alpha$ anomalies (an
idealized first-order phase transition would yield a jump in
$\Delta L$ and a $\delta$ peak in $\alpha$). Despite the
different shapes of the $\alpha$ anomalies of LaCoO$_3$ and of
\PEC, the total length changes $\Delta L/L$ up to 150~K are in
the range of $2.3\cdot 10^{-3}$ to $3.8\cdot 10^{-3}$ for all
the samples shown in the right panel of \fref{alpha}.

\begin{figure*}[!tb]
\includegraphics{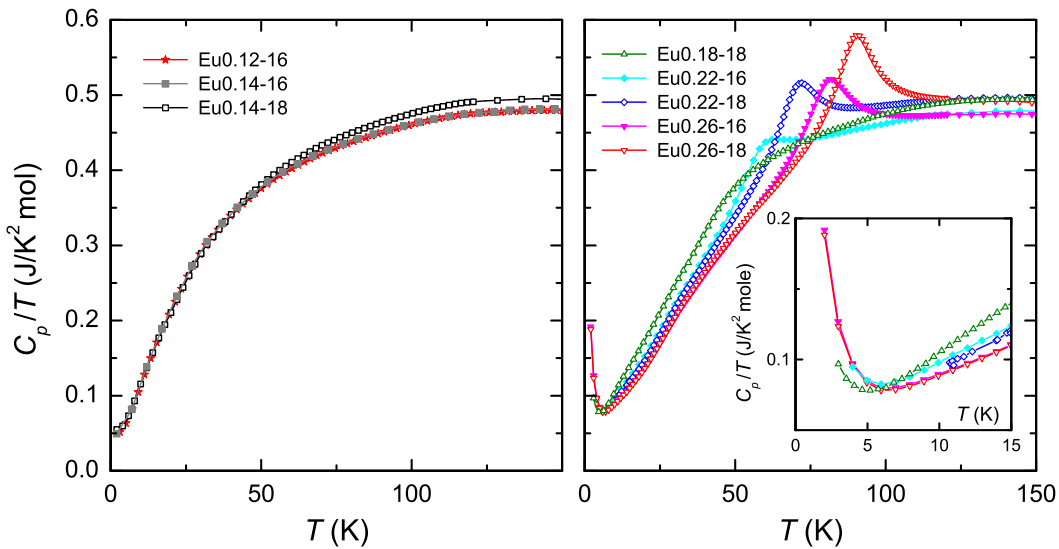}
\caption{(Color online)\label{Cp} The temperature dependence of
the specific heat divided by temperature of \PEC. The left
panel shows the smooth ferromagnetic crossover of the LD (less
Eu) samples. In the right panel, the spin-state transition
anomaly is observed for the HD (more Eu) samples. The inset of
the right panel shows an additional low-temperature anomaly,
which is related to a spin ordering of the \coiiii\ ions below
5~K. This anomaly is absent for the LD samples.}
\end{figure*}

In \fref{Cp}, the specific heat measurements are illustrated.
Again, we observe the clear difference between the LD and HD
samples. The HD samples show pronounced anomalies at
temperatures, which well agree with those of the corresponding
anomalies in $\alpha$. The shape of these anomalies is also
rather symmetric as expected for a broadened first-order phase
transition. In addition, there are low-temperature upturns in
$C_p/T$ for all HD samples measured to low enough $T$ (see
inset), which indicate the occurrence of another phase
transition at $T < 2$~K. Both features are absent in the LD
samples as shown in the left panel of \fref{Cp}. Another aspect
is seen in the high-temperature limit: independent of the
presence or absence of the anomalies, all $C_p/T$ curves of the
\osix\ samples practically meet in a single line, which is
slightly lower than the corresponding line where all the
$C_p/T$ curves of the \oeight\ samples meet. This difference
arises from the lowering of the frequencies of those phonons,
which contain vibrations of heavy oxygen. Thus, at fixed $T$,
the number of excited oxygen modes in an \osix\ sample is less
than in the corresponding \oeight\ sample and therefore the
lattice specific heat is less in the former. This systematic
behavior of our samples clearly confirms the high reliability
of the oxygen exchange in this series. Similar isotope effects
in $C_p$ of binary solids ZnO and PbS were theoretically and
experimentally studied in Refs.~\onlinecite{serrano.prb06.ZnO,
cardona.prb07.PbS}.

\begin{figure}[!tb]
\includegraphics{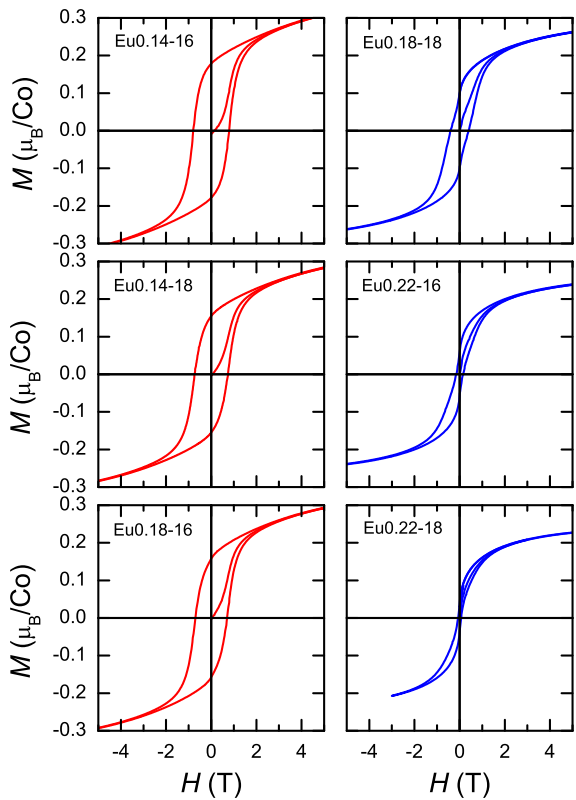}
\caption{(Color online)\label{loops-2K} Magnetization of \PEC\
at 2~K as a function of magnetic field. Magnetization was
measured after zero-field cooling. The paramagnetic
contribution of Pr$^{3+}$ is subtracted, based on the
measurements of PrCoO$_3$ weighted by the actual Pr content in
each sample (see also \fref{loops-14T}). The left panels
correspond to the LD samples, the right panels are for the HD
samples.}
\end{figure}

\begin{figure}[!tb]
\includegraphics{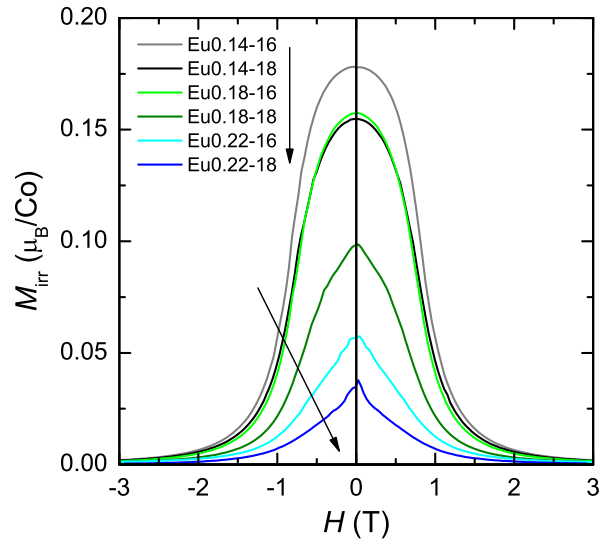}
\caption{(Color online)\label{M-irrev} Irreversible part of the
magnetization of \PEC\ as a function of magnetic field.
$M_\mathrm{irr}$ is defined as the half-difference of the
magnetization measured in increasing and decreasing magnetic
fields. The arrows indicate the sequence of curves (from LD to
HD samples). }
\end{figure}

Now, we turn to the magnetization measurements.
Figure~\ref{loops-2K} displays representative $M(H)$ curves
measured at 2~K on LD (left panels) and HD samples (right
panels). Again, there is a clear difference; the LD samples
exhibit a pronounced hysteresis, as it is typical for
ferromagnets, with a large remanence and a coercitivity of the
order of 1~T. A hysteretic behavior is also present for the HD
samples, but it is much weaker and also qualitatively
different. This is best seen by considering the irreversible
parts of the $M(H)$ curves. The irreversible part of the
magnetization $M_\mathrm{irr}$ is defined as the
half-difference of the $M(H)$ curves measured with increasing
and decreasing magnetic field. In this consideration, the
virgin curve was withdrawn. The result is shown in
\fref{M-irrev}. The LD samples have broad $M_\mathrm{irr}(H)$
curves with a weak negative curvature around zero field, while
the $M_\mathrm{irr}(H)$ curves of the HD samples have
essentially an almost cusp-like shape centered at zero field.

Such a crossover from the high-remanence state to the soft
magnetic state was observed both for $A$-site isovalent
substitution \cite{paraskevopoulos.prb01} and on heterovalent
doping with Sr.\cite{aarbogh.prb06} In contrast to our case,
the results of Ref.~\onlinecite{paraskevopoulos.prb01} for
$R_{0.67}$Sr$_{0.33}$CoO$_3$ demonstrated that the coercitivity
{\em increases} significantly with decreasing the mean radius
of the $R$ ion. In our observations, the coercitivity {\em
decreases} with decreasing ionic radius (increasing Eu
content). In the case of La$_{1-x}$Sr$_x$CoO$_3$,
\cite{aarbogh.prb06} the coercitivity peaks at $x \approx 0.12$
and drops almost to zero at $x = 0.18$, just before the
transition to the ferromagnetic metallic state. This effect was
suggested to originate from the formation of thermally stable
nanoscale ferromagnetic droplets in a nonmagnetic matrix that
results in the peak in coercitivity. With further increase of
the Sr content, these droplets form the multi-domain clusters,
which percolate at the crossover to the FM state at $x = 0.18$.
In our case, the coercitivity grows from about zero for $y =
0.26$ to 8~kOe for $y = 0.12$, with the steep increase being
visible at $y \approx y_\mathrm{th}$, and retains the same
value up to the end-member
Pr$_{0.7}$Ca$_{0.3}$CoO$_3$.\cite{tsubouchi.prb04} This
suggests that the nanosize ferromagnetic particles nucleate at
$y \approx y_\mathrm{th}$ and then grow up to $y = 0$ without a
crossover to the multidomain-droplet state.

The occurrence of ferromagnetic-like hysteresis loops in the LD
samples is rather surprising because on increasing temperature
the $C_p(T)$ and $\alpha(T)$ data do not exhibit any clear
indications of a well-defined ordering transition (see
\fref{alpha} and \fref{Cp}). The same is true for temperature
dependent measurements of the d.c. magnetization in finite
magnetic fields (not shown). Thus, we think that the origin of
the hysteresis loops is most probably related to the
coexistence of different ``phases'' with competing magnetic
interactions and, simply speaking, the hysteresis measures the
``ferromagnetic'' phase fraction. In such a case, there is no
real long-range order and one may expect a rather continuous
crossover behavior on decreasing temperature. In order to
estimate a characteristic crossover temperature, we considered
the vanishing of the remanent magnetization $M_R(T)$ in zero
field [after having performed a full hysteresis loop
$M(H,T=2\mathrm K)$] as a function of increasing temperature.
Typically, $M_R(T)$ vanishes around 40 to 50~K with a weak
dependence on $y$ and on the oxygen isotope, which will be
discussed below.

\begin{figure}[!tb]
\includegraphics{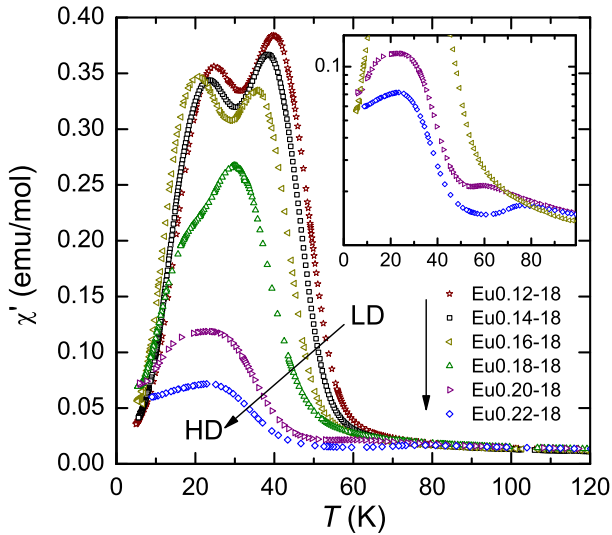}
\caption{(Color online)\label{chi} Temperature dependence of
the real part of the a.c.\ magnetic susceptibility of \PEC. The
arrows indicate the sequence of curves (from LD to HD samples).
In the inset, the spin-state transition region is enlarged (the
units are the same as in the main plot; note the logarithmic
scale along the vertical axis.).}
\end{figure}

In addition, we measured the real part of the a.c. magnetic
susceptibility $\chi'$. As shown in \fref{chi}, the LD samples
have a strong increase of $\chi'$ at about 50~K, which is
followed by a complicated two-hump feature on further cooling.
With increasing Eu content this increase systematically shifts
toward lower temperature and we used the maximum slope of
$\chi'(T)$ as the criterion for the definition of the crossover
temperature for the phase diagram. For the HD samples, $\chi'$
shows a much weaker increase and only one broad hump on further
cooling. Another difference is shown in the inset of
\fref{chi}; the HD samples show a stepwise increase in
$\chi'(T)$, which does not occur in the LD samples.

\begin{figure}[!tb]
\includegraphics{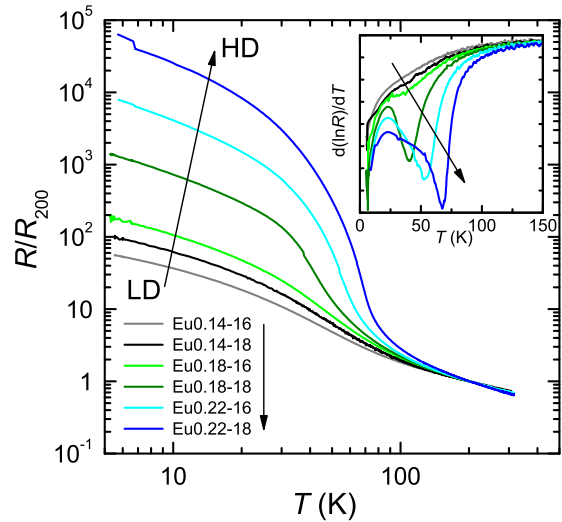}
\caption{(Color online)\label{r(t)-log} Normalized electrical
resistivity of \PEC\ as a function of temperature. Arrows
indicate the sequence of curve (from LD to HD samples). The
inset shows the logarithmic derivatives of $R$ (in arbitrary
units) as a function of temperature. The spin-state-transition
related peak is clearly present for the HD samples only.}
\end{figure}

This transition is not only seen in the thermodynamic
properties ($\alpha$, $C_p$, and $\chi$) but is also visible in
the electrical resistivity. The room temperature resistivity is
approximately the same for all the samples (of about $10^{-3}
\Omega$\thinspace cm) and the data presented in Fig.
\ref{r(t)-log} demonstrate that it has a weak semiconducting
behavior. Below 100~K, all the HD samples show a transition to
a high-resistive state, which does not occur in the LD samples.
This is most clearly visible in the temperature dependence of
the logarithmic derivative of the resistivity (see inset of
\fref{r(t)-log}). The corresponding transition temperatures
coincide well with the temperatures obtained in the thermal
expansion and specific heat (\fref{alpha} and \fref{Cp}).

\section{Discussion}

Summarizing the data presented above, we observe in all HD
samples clear signatures of a first-order phase transition,
which on increasing temperature causes (i) a pronounced lattice
expansion, (ii) a gain in entropy, (iii) a step-like increase
of the magnetization, and (iv) a drop in the electrical
resistivity. All these features can be naturally explained by a
spin-state transition from a LS to a higher-spin state of the
\coiii\ ions, because the latter have (i) a larger volume, (ii)
more spin entropy, and (iii) a finite magnetic moment. To be
more precise, the larger unit cell volume, which is discussed
here, does not inevitably mean an increase of a \coiii\ ionic
radius. It is also possible that as shown in
Ref.~[\onlinecite{tong.jpsj09}], the increase of the Co--O--Co
bond angle (and concomitant CoO$_6$ octahedra distortions)
could solely be responsible for the anomalous volume expansion
upon the SST.

The drop in the resistivity (iv) suggests that the higher spin
state is the IS state, because LS-\coiiii/IS-\coiii\ neighbors
allow for an easy electron transfer via the \eg\ orbitals,
whereas for LS-\coiiii/LS-\coiii\ only a hopping via the \tg\
orbitals, which have much less overlap, is possible and for the
LS \coiiii/HS \coiii\ combination the electron transfer is
suppressed by the so-called spin-blockade
effect.\cite{maignan.prl04} The other possible scenario, which
could have explained the observed features, would be a charge
ordering in the low-temperature phase. However, no superlattice
reflections (which would be a fingerprint of charge ordering)
in NPD and XRD data \cite{tong.jpsj09, tsubouchi.prb02} were
observed in the closely related Pr$_{0.5}$Ca$_{0.5}$CoO$_3$
compound. So, based on the neutron and X-ray diffraction data
we can refute this alternative scenario.

The observed transition to a LS state with decreasing
temperature exists only for the HD samples and is absent in the
LD samples. That means that in the LD samples the \coiii\ ions
remain in the higher-spin state down to the lowest temperature,
and the presence of LS-\coiiii/IS-\coiii\ neighbors appears to
be most probable, because the \eg\ electron transfer favors a
ferromagnetic alignment of the \tg\ moments via the
double-exchange mechanism. This would be a natural source of
the almost ferromagnetic low-temperature $M(H)$ hysteresis
loops in the LD samples. The fact that we are not dealing with
a real long-range ferromagnetic order manifests itself in a
more insulating behavior of the \PEC\ series as compared, e.g.,
to La$_{0.75}$Ba$_{0.25}$CoO$_3$ or
La$_{0.75}$Sr$_{0.25}$CoO$_3$. The latter are metallic in the
entire temperature range and show ferromagnetic ordering
transitions at about 220~K, which is most probably triggered by
the double-exchange mechanism.\cite{kriener.prb04}

To get more information about the actual low-temperature spin
states, let us consider the low-temperature high-field
magnetization. For the above-mentioned ferromagnetic metals
La$_{0.75}$(Ba,Sr)$_{0.25}$CoO$_3$, a saturation magnetization
of about 1.7\muBco\ is observed, which is close to the expected
value of 1.75\muBco\ corresponding to a 1:3 ratio of
LS-\coiiii/IS-\coiii. For the actual series, one has to
consider also the (Pr$_{1-y}$Eu$_y$) magnetism of the $4f$
shell. According to Hund's rules, Eu$^{3+}$ ions have a
nonmagnetic $^7$F$_0$ ground state, but in finite magnetic
fields this state is mixed with the higher-lying $J = 1, 2,
\dots$ multiplets that results in a finite van Vleck
susceptibility. The ground state of free Pr$^{3+}$ ions is a
$^3$H$_4$ multiplet with a total moment $J=4$, which splits
into 9 singlets in the orthorhombic crystal field, and again
these different singlets are mixed in a finite magnetic field.
As a consequence, one may expect a considerable low-temperature
magnetization $M_\mathrm{RE}(H)$ from (Pr$_{1-y}$Eu$_y$), which
is essentially linear in field. The detailed form and the exact
absolute value of this $M_\mathrm{RE}(H)$ depend on details of
the crystal field, which are not known for our samples, but one
can expect that the crystal fields here will not differ too
much from those in EuCoO$_3$ and PrCoO$_3$. Since in these
undoped compounds the \coiii\ ions are in the LS state, the
contributions of Eu$^{3+}$ and Pr$^{3+}$ are directly measured
by the low-temperature $M(H)$ curve (apart from additional
small contributions due to impurities or oxygen
off-stoichiometry). In previous studies on EuCoO$_3$ and
PrCoO$_3$,\cite{berggold.prb08, baier.prb05} we found
$M(H=14\mathrm T)$ values of about 0.17\muB\ and 0.55\muB\ per
formula unit, respectively.

\begin{figure}[!tb]
\includegraphics{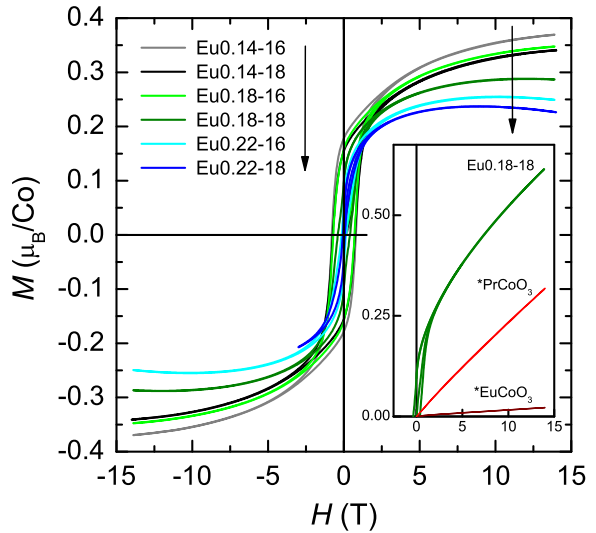}
\caption{(Color online)\label{loops-14T} Zero-field-cooled
magnetization of \PEC\ at 2~K versus magnetic field. Arrows
indicate the sequence of curve (from LD to HD samples). The
paramagnetic contribution of the Pr ions is subtracted, based
on the measurements of PrCoO$_3$. The inset shows (the units
are the same as in the main plot) the magnetization of \PEC\
with $y=0.18$ and \oeight\ before the subtraction of the Pr
contribution. The data for PrCoO$_3$ and EuCoO$_3$ normalized
to the respective actual ionic contents of Pr and Eu in this
sample are shown for comparison (see the text for details).}
\end{figure}

To estimate the Co contributions in the \PEC\ samples, we
therefore subtracted the $M(H)$ curve of PrCoO$_3$ weighed by
the respective Pr content from the measured $M(H)$ curves. The
resultant \mREonH{Co}\ curves are summarized in
\fref{loops-14T} and, at least for the samples with higher Eu
contents, there seems to be some ``overcorrection''. Since the
Pr contribution \mREonH{Pr}\ is already somewhat overestimated,
we did not perform an additional subtraction of an Eu
contribution \mREonH{Eu}. Note that due to the different Eu and
Pr contents, \mREonH{Eu}\ is only of about 15\% of \mREonH{Pr}\
in all samples (see inset of \fref{loops-14T}) and thus, it is
within the uncertainty of the estimates of the entire
contribution of (Pr$_{1-y}$Eu$_y$). Independent of this
uncertainty, it is clearly seen that the \mREonH{Co}\ curves
tend to the saturation value of about 0.3\muBco. This value
agrees well with that of all \coiiii\ ions in the LS state,
i.e., 1\muB/\coiiii, with all the \coiii\ ions being in the
nonmagnetic LS state. For the HD samples such a state appears
plausible, because our data yield clear evidence for a SST from
the IS to the LS state of \coiii\ ions on decreasing
temperature. For the LD samples, however, the LS state of the
\coiii\ ions is very unlikely because in such a case, both the
LD and HD samples would have the same spin state combination
(LS \coiii/LS \coiiii) at low $T$ and thus one could not
explain the qualitatively different $M(H)$ loops (see
\fref{loops-2K} and \fref{M-irrev}). As discussed above, the LD
samples are closer to a ferromagnetic behavior than the HD
samples, which agrees with the expectations that due the lower
chemical pressure in the LD samples at least more (maybe all)
\coiii\ ions remain in the IS state. The fact that,
nevertheless, the magnetization hardly exceeds that of the HD
samples gives further evidence that there are competing
magnetic interactions, e.g., AFM \coiii\ -- \coiii\ and FM
\coiii\ -- \coiiii, which prevent a real long-range magnetic
order in the LD samples. The presence of AFM clusters could
also explain the finite high-field slope of the $M(H)$ curves
of the LD samples, although this argument has to be treated
with some caution because of the uncertainty in subtracting the
background $M_\mathrm{RE}(H)$.

\begin{figure}[!tb]
\includegraphics{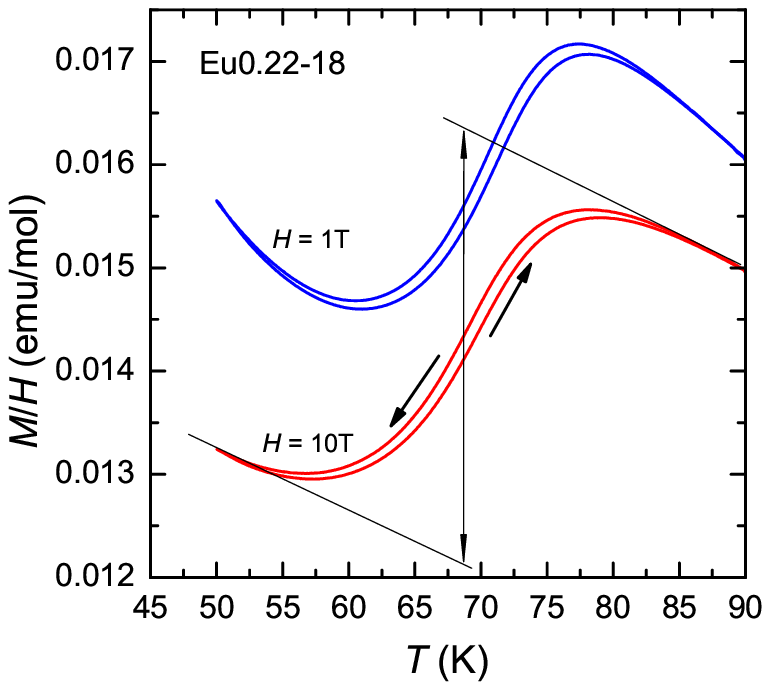}
\caption{(Color online)\label{sst-shift} Temperature dependence
of the static magnetic susceptibility of \PEC\ with $y = 0.22$
and \oeight\ in the vicinity of the spin-state transition at $H
= 1$ and 10~T. The rate of temperature change was $\pm
0.6$~K/min. Bold arrows indicate the sequence of the
magnetization changes, confirming the first order of the
transition. The long arrow and straight lines show
schematically the procedure to determine $\Delta M$ at the
transition.}
\end{figure}

Next, we will discuss whether the SST of the HD samples is
complete, meaning which amount of the \coiii\ ions is involved
in the SST. It is straightforward to analyze this via the
entropy change at the SST, which can be obtained either
directly from the specific heat data or by using the
Clausius-Clapeyron equation $d T_\mathrm{SS}/d B = -\Delta
M/\Delta S_\mathrm{magn}$, which, for a first-order phase
transition, relates the field dependence of the transition
temperature $T_\mathrm{SS}$ to the ratio of the discontinuous
changes $\Delta M$ and $\Delta S_\mathrm{magn}$ of the
magnetization and the magnetic entropy, respectively. In
\fref{sst-shift} we compare $M(T)$ curves for fields of 1 and
10~T, each of them measured both on heating and cooling. A
small but finite hysteresis is visible in both curves and it is
also seen that the transition shifts to lower $T$ with
increasing field. For the quantitative analysis we linearly
extrapolated the measured $M(T)$ curves well above and below
$T_\mathrm{SS}$ and found $\Delta M$ as a step change between
these lines, as indicated by the straight lines and the thin
arrow in \fref{sst-shift}. With $d T_\mathrm{SS}/d B =
0.15$~K/T and $\Delta M = 410$~emu/mol obtained in this way, we
estimate the magnetic entropy change $\Delta S_\mathrm{magn} =
2.7$\JKm.

\begin{figure}[!tb]
\includegraphics{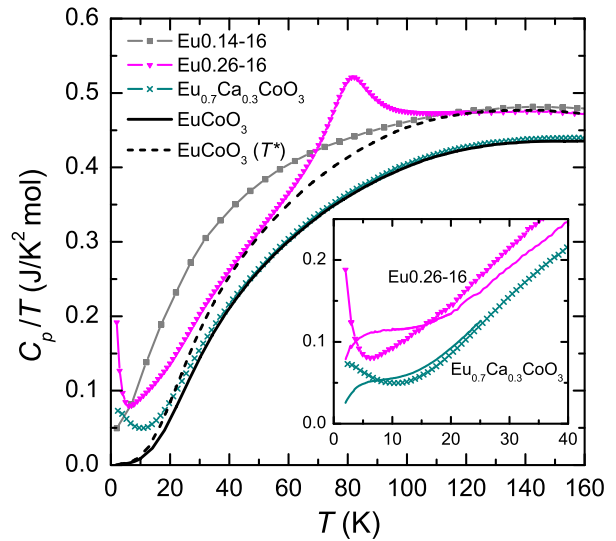}
\caption{(Color online)\label{Cp-backgr} Temperature dependence
of the specific heat divided by temperature of \PEC\ ($y =
0.14, 0.26$). The data for Eu$_{0.7}$Ca$_{0.3}$CoO$_3$ and
EuCoO$_3$ are also shown. The dashed line represents the data
for EuCoO$_3$ after scaling the temperature by $T^* = 0.92 T$
(see text for details). The inset demonstrates the effect of a
magnetic field on the low-temperature anomaly in $C_p/T$;
curves with symbols are for $H=0$ (the same as in the main
plot) and lines are for $H=10$~T.}
\end{figure}

In order to analyze the magnetic specific heat, we present in
\fref{Cp-backgr} the $C_p/T$ plots of two members of the \PEC\
series, which are characteristic either for the LD or the HD
samples; see \fref{Cp}. Obviously, the $C_p/T$ curves of the LD
and HD samples coincide above about 120~K, while for lower $T$
systematic differences are present: there are two anomalies for
the HD samples, in contrast to one very broad shoulder for the
LD samples. Interestingly, the total entropy change from 2 to
120~K of the HD and LD samples differs only by about 2\JKm\
and, because of the low-temperature anomaly, this difference
would further decrease if the measurements were performed to
lower $T$. One may even speculate that finally this difference
should more or less completely vanish for $T \rightarrow 0$~ K.
This could be interpreted in the following simple picture: for
both, the LD and the HD samples (i) the phonon specific heat is
essentially the same, and (ii) at high $T$ the total magnetic
entropy is given by $R(0.3 \ln 2 + 0.7 \ln 3)$. Here, $R =
8.31$\JKm\ is the gas constant and $\ln 2$ and $\ln 3$ stem
from $\ln(2S+1)$ with $S=1/2$ for LS \coiiii\ and $S=1$ for IS
\coiii, respectively. On decreasing $T$, this magnetic entropy
continuously becomes frozen for the LD samples without showing
a well-defined magnetic phase transition because of the
presence of competing magnetic interactions. For the HD
samples, in contrast, a part of the magnetic entropy freezes at
about 80~K due to a SST of \coiii\ ions and the rest via an
additional low-$T$ transition.

To get more information about the origin of the low-$T$
transition, we also measured $C_p$ of EuCoO$_3$ and
Eu$_{0.7}$Ca$_{0.3}$CoO$_3$. For EuCoO$_3$, the \coiii\ ions
are known to be in the LS state up to above room
temperature.\cite{baier.prb05} The $C_p/T$ data of both Eu
compounds superimpose on each other everywhere except in the
low-temperature region. It appears natural that (because of the
very small Eu ions) all the \coiii\ ions remain in the LS state
even in the Ca-doped compound. So, the difference below 20~K
and the upturn at low temperatures should be merely related to
the \coiiii\ ions. In addition, we also found that a magnetic
field of 10~T strongly suppresses the low-temperature upturn
(see inset of Fig. \ref{Cp-backgr}) confirming that the extra
entropy is of a magnetic nature. Moreover, the entropy
difference of 1.2\JKm\ between EuCoO$_3$ and
Eu$_{0.7}$Ca$_{0.3}$CoO$_3$ in the temperature range down to
2~K is quite comparable to the full magnetic entropy $\Delta
S_\mathrm{Co4+} = 0.3 R \ln 2 = 1.73$\JKm\ of \coiiii\ ions.

Similar low-temperature anomalies in $C_p(T)$ were observed in
closely related Pr$_{1-x}$Ca$_x$CoO$_3$
[\onlinecite{tsubouchi.prb04}] and in
(Pr$_{1-y}$Sm$_y$)$_{1-x}$Ca$_x$CoO$_3$
[\onlinecite{fujita.jpsj05}] compounds as well as in undoped
LaCoO$_3$ [\onlinecite{he.apl09}]. C.\ He and co-workers
\cite{he.apl09} have interpreted the observed anomaly as a
Schottky anomaly associated with the first excited (by 0.6~meV)
spin state of the \coiii\ ion. Despite the similar magnetic
field dependence of the anomaly that was observed in
Ref.~\onlinecite{he.apl09} and in our data, the origin of  this
effect must be different. In our case, the anomaly appears only
for the HD samples and, if the energy gap would arise from the
proposed Schottky anomaly, it should increase with increasing
distortion that would result in a suppression of the anomaly.
But this is not the case: the anomaly is essentially the same
for all HD sample. (These curves are not shown in
\fref{Cp-backgr}, because they are almost coincide). Moreover,
our samples contain a significant fraction ($\approx 30\%$) of
magnetic \coiiii\ ions, in stark contrast to slightly
oxygen-deficient LaCoO$_3$. Thus, we conclude that the
low-temperature upturns in $C_p/T$ observed in
Eu$_{0.7}$Ca$_{0.3}$CoO$_3$ and in the HD samples of the \PEC\
series provide an evidence for some kind of magnetic ordering
of a (dilute) system of \coiiii\ ions in a background of
nonmagnetic LS \coiii\ ions.

A quantitative determination of the entropy change related to
the SST requires the knowledge of the phonon contribution. To
estimate this contribution, we use the $C_p$ data of EuCoO$_3$
and rescale the temperature axis by $T^* = 0.92 T$. This
procedure does not have a {\em quantitative} meaning, but the
obtained $C_p(T^*)/T^*$ curve matches the data of the HD
samples of the \PEC\ series both on the high-temperature tail
and in the region below the SST by using a single correction
coefficient only. Thus, we consider this as a reasonable phonon
background and its subtraction from the measurements on the HD
samples yields entropy changes at the SST ranging from 2.5 to
2.6\JKm\ for $y=0.22$ to 0.26. These values are close to
$\Delta S_\mathrm{magn}$ obtained independently from the
observed field dependence of $T_\mathrm{SS}$ via the
Clausius-Clapeyron equation. This entropy change is
significantly smaller than the expected $\Delta
S_\mathrm{LS-IS} = 0.7 R \ln (2S+1) = 6.4$\JKm, when all
\coiii\ ions would be transformed from the LS to the IS state.
Interestingly, the obtained entropy change is much closer to
$0.3 R \ln(2S+1) = 2.74$\JKm, suggesting that on average each
\coiiii\ ion induces a SST in only one (probably) neighboring
\coiii\ ion. This is very different from the results for very
lightly Sr-doped LaCoO$_3$,\cite{podlesnyak.prl08,
yamaguchi.prb96} where each Sr (or \coiiii\ thus created)
promotes in average six neighboring \coiii\ to the IS state.
Thus, apparently the extent, to which the LS \coiii\ ions are
promoted to the IS state depend on details of the system. We
suspect that this very different behavior of lightly Sr-doped
LaCoO$_3$ and the \PEC\ series arises from two effects: (i) a
larger crystal-field splitting and (ii) a reduced electron
hopping because the $e_g$ bandwidth is reduced in the stronger
distorted structure for the smaller rare-earth ions
[(Pr$_{1-y}$Eu$_{y}$) instead of (La$_{1-x}$Sr$_{x}$)]. Note
that in the 50\%-doped Pr$_{0.5}$Ca$_{0.5}$CoO$_3$ compound
\cite{tsubouchi.prb02} the spin-state-transition entropy was
estimated to 4.7\JKm, which is equal to the entropy if all 0.5
\coiii\ are promoted from the LS to IS state. This does,
however, not contradict our finding, because the half-doped
case contains equal amounts of \coiii\ and \coiiii\ ions.

\section{Phase diagram}

\begin{figure}[!tb]
\includegraphics{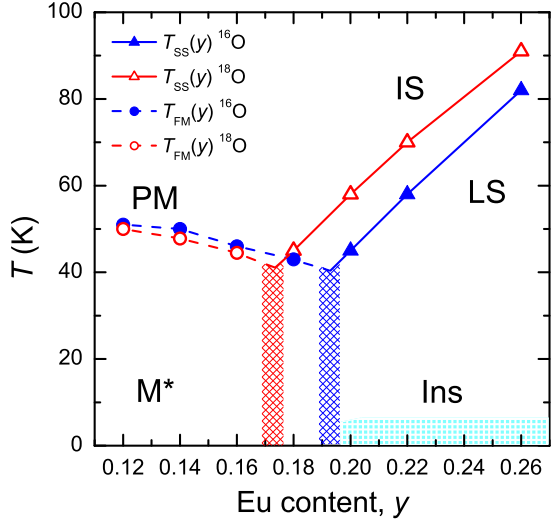}
\caption{(Color online)\label{diagram} Magnetic/spin-state
phase diagram of \PEC. The symbols show the corresponding
transition temperatures as a function of the Eu doping $y$:
triangles denote the spin-state transition temperatures and
circles are for the magnetic crossover temperature. Lines are
guides for the eye. `PM' denotes the paramagnetic metallic
region, `M$^*$' the ``bad metal'' area, `Ins' means insulating
state; `LS' and `IS' denote the low-spin and intermediate-spin
states of \coiii, respectively. The shaded area for the low
temperatures and high Eu contents schematically shows the
low-temperature magnetic ordering of the \coiiii\ ions, which
occurs for the same $y$ as the spin-state transition of the
\coiii\ ions.}
\end{figure}

Based on all obtained results, we derive the phase diagram
describing magnetic and spin-state transitions (see
\fref{diagram}). There are three magnetic/spin states of
\coiii\ ions in the diagram. (Because we are not aware of any
compound where \coiiii\ realizes another spin state than the LS
state with $S = 1/2$, only the spin state of \coiii\ ions is
discussed.) For $y < y_\mathrm{th}$ (LD samples), the ground
state becomes ferromagnetic with a large remanence and
coercitivity. A fuzzy $T_\mathrm{FM}(y)$ line separates it from
the high-temperature paramagnetic state. For Eu contents just
below $y_\mathrm{th}$, the low-temperature saturation
magnetization exceeds the maximum value for the LS state of all
cobalt ions. This implies that in the absence of the FM
ordering (due to LS-\coiiii/IS-\coiii\ interaction), the
\coiii\ ions just below $y_\mathrm{th}$ would still show a
transition to the LS state. Moreover, there is no
low-temperature upturn in the $C_p(T)/T$ data for the LD
region, and a magnetic field of 10~T does not affect the
specific heat at all. This suggests that the \coiiii\ ions are
already ordered at high temperatures giving rise to the
coercitivity. The low-temperature resistivity in these LD
samples is lower by several orders of magnitude compared to
that of the HD samples, but there is no real metallicity in the
$R(T)$ dependence. The saturation magnetization reaches only
0.35\muB\ per Co site (up to 0.5\muBco\ for the end-member
Pr$_{0.7}$Ca$_{0.3}$CoO$_3$ with $T_{FM}\approx 50$~K
[\onlinecite{tsubouchi.prb04}]). A fully polarized \coiii\
sublattice in the IS state should result in a much higher
magnetization of 1.4\muBco\ (and even more so for HS \coiii\
ions). The \tgeg{5}{1} intermediate-spin state makes the double
exchange mechanism \cite{zener.pr51} possible due to the
presence of the itinerant \eg\ electrons that can explain the
simultaneous nucleation of ferromagnetism and conductivity. We
can qualitatively interpret these data (fuzzy transition, low
magnetic moment, absence of a real metallic behavior) in the
picture of a strongly inhomogeneous (phase-separated) state
with ferromagnetic metallic clusters embedded into a
nonmagnetic and insulating (or, at least, less magnetic and
less conducting) background.

With increasing temperature the LD samples show a smooth
crossover to the paramagnetic state, which is observable as a
broad hump in the temperature dependence of the specific heat;
see Figs.~\ref{Cp} and \ref{Cp-backgr}. The corresponding
entropy associated with the crossover for Eu$_{0.14}$-16 is
estimated to 6.7\JKm, which is smaller than the total magnetic
entropy $\Delta S_\mathrm{FM}=0.3 R \ln 2 + 0.7 R \ln 3 =
8.1$\JKm\ assuming that the \coiiii\ ions are in LS ($S = 1/2$)
and all \coiii\ ions are in the IS state ($S = 1$).

With increasing Eu content, i.e.\ increasing chemical pressure,
the \coiii\ LS state becomes more and more stabilized and above
a certain critical value, temperature-dependent spin-state
transitions occur for the HD samples. However, as discussed in
the previous section, in average every \coiiii\ ion induces
this spin-state transition in only one neighboring \coiii\ ion,
i.e., less than half of the \coiii\ ions actually show the
spin-state transition.

The isotope effect is clearly observed for the spin-state
transition $T_\mathrm{SS}$. The temperature $T_\mathrm{FM}$ has
only marginal but visible dependence on the isotope content.
However for both phase boundaries [$T_\mathrm{SS}(y)$ and
$T_\mathrm{FM}(y)$], the isotope exchange is equivalent to the
change in $y$ by approximately 0.02. This means that the static
distortions due to the change in the mean radius of rare-earth
ions are somehow equivalent to changes in the lattice dynamics
due to the isotope exchange, although the physical mechanisms
are apparently different. An increase of the Eu content leads
mainly to an increase of the \tg\ -- \eg\ crystal field
splitting, which stabilizes the LS state. On the other hand,
the main effect of the oxygen isotope substitution is a change
of the effective intersite hopping, i.e.\ of the bandwidth. For
the LD samples both phases, FM and PM are ``bad metallic'' with
magnetic \coiii\ ions and partially filled \eg\ bands. The
bandwidth changes with the oxygen isotope substitution but the
resulting effect on $T_\mathrm{FM}$ is relatively
weak.\cite{babushkina.jap98} At the same time, the SST of the
HD samples (in the right part of the phase diagram) is
accompanied by a more drastic change of the electronic
structure. Whereas in the insulating LS state the \eg\ bands
are practically empty, they are partially occupied in the
high-temperature phase with the promotion of a part of the
\coiii\ ions to the IS state. In this case, the narrower \eg\
bands for the \oeight\ samples start to overlap with the \tg\
levels much later than for the \osix\ samples. Consequently,
the \eg\ bands for the \oeight\ samples become occupied later
and are filled more slowly. This strongly stabilizes the LS
state and shifts the SST upwards in the \oeight\ case. Close to
the LS -- IS crossover even small changes in hopping $t$ and
bandwidth $W \sim 2zt$ ($z$ is the number of nearest neighbors)
lead to a pronounced shift of the phase
equilibrium.\cite{babushkina.jap98} Indeed, we see that
$T_\mathrm{SS}$ is much higher for the samples with the heavier
oxygen isotope. Note, however, that the isotope effect for
$T_\mathrm{FM}$, being much weaker, is of the opposite sign.
($T_\mathrm{FM}$ is slightly lower for the \oeight\ samples
with the narrower \eg\ bands.) This fits to our expectations,
because the ferromagnetism of the low-Eu doped samples (with
metallic conductivity) should be stabilized by the
double-exchange mechanism according to which $T_\mathrm{FM}$ is
proportional to the effective bandwidth of the itinerant
electrons. Itinerant electrons are less affected by the lattice
and hence less sensitive to the isotope composition, although
the bandwidth can be renormalized due to the electron-phonon
interaction. However, such effects are usually rather small
since the dimensionless electron-phonon coupling constant does
not depend explicitly on the atomic mass, if the system does
not correspond to the regime of small polarons. The situation
here has some similarities with the isotope effect in
manganites with competing states of a charge-ordered insulator
and ferromagnetic metal.\cite{babushkina.prb08}

\section{Summary}

The magnetic/spin-state phase diagram of \PEC\ series was
obtained on the basis on the measurements of the specific heat,
thermal expansion, magnetization and resistivity. The phase
diagram reveals three different states depending on the static
distortions (Eu content), the oxygen-isotope mass, and the
temperature. The samples with the lower Eu concentrations are
ferromagnetically ordered up to moderate temperatures (about
50~K) most probably due to the \coiiii\ (LS, $S = 1/2$) --
\coiii\ (IS, $S = 1$) interaction of the double-exchange type,
with the promotion of the \coiii\ ions to the higher-spin
state. As the Eu doping increases, the \coiii\ LS ($S = 0$)
state becomes stabilized and the magnetic ordering of the
\coiiii\ ions is suppressed to temperatures well below 5~K. At
higher temperatures, we observe a first-order spin-state
transition from the LS to the IS state of \coiii, which is
accompanied by a strong decrease of the electrical resistivity.
Again this temperature-activated spin-state transition is
promoted by the $e_g$ electron hopping to the neighboring
\coiii\ ions, but on average every \coiiii\ ions induces a
spin-state transition in only one \coiii\ ion.

The oxygen-isotope exchange (\osix\ to \oeight) shifts the
phase boundaries to the lower Eu concentration, i.e., an
increase of the oxygen mass acts similarly to an increase of
the Eu content. Nevertheless, the mechanisms of such shifts
seem to be different: increasing the Eu content mainly
increases the crystal field splitting whereas the main effect
of the oxygen-isotope substitution from \osix\ to \oeight\ is a
decrease of the effective band width, but both effects favor
the stabilization of the insulating state with LS \coiii. Note,
however, that for a given composition, the isotope effect on
the spin-state transition (in the samples with high Eu content)
is quite strong, whereas it is much weaker and of opposite sign
for the magnetic transitions in the samples with the low Eu
contents.

We expect that the regularities observed in these systems
should also be applicable to other cobaltites with spin-state
transitions. For example, the observed correlation of the
\coiiii-related $C_p/T$ upturn at low temperatures with the
presence of the spin-state transition at a higher temperature
can be seen also in the Pr$_{1-x}$Ca$_x$CoO$_3$
[\onlinecite{tsubouchi.prb04}] and in
(Pr$_{1-y}$Sm$_y$)$_{1-x}$Ca$_x$CoO$_3$
[\onlinecite{fujita.jpsj05}] series.

\begin{acknowledgments}
The present work was supported by the Russian Foundation for
Basic Research (projects 07-02-00681, 07-02-91567 and
10-02-00598), and by the Deutsche Forschungsgemeinshaft via SFB
608 and the German-Russian project 436 RUS 113/942/0.
\end{acknowledgments}


\begin{thebibliography}{51}
\expandafter\ifx\csname
natexlab\endcsname\relax\def\natexlab#1{#1}\fi
\expandafter\ifx\csname bibnamefont\endcsname\relax
  \def\bibnamefont#1{#1}\fi
\expandafter\ifx\csname bibfnamefont\endcsname\relax
  \def\bibfnamefont#1{#1}\fi
\expandafter\ifx\csname citenamefont\endcsname\relax
  \def\citenamefont#1{#1}\fi
\expandafter\ifx\csname url\endcsname\relax
  \def\url#1{\texttt{#1}}\fi
\expandafter\ifx\csname
urlprefix\endcsname\relax\def\urlprefix{URL }\fi
\providecommand{\bibinfo}[2]{#2}
\providecommand{\eprint}[2][]{\url{#2}}

\bibitem[{\citenamefont{Imada et~al.}(1998)\citenamefont{Imada, Fujimori, and
  Tokura}}]{imada.rmp98}
\bibinfo{author}{\bibfnamefont{M.}~\bibnamefont{Imada}},
  \bibinfo{author}{\bibfnamefont{A.}~\bibnamefont{Fujimori}}, \bibnamefont{and}
  \bibinfo{author}{\bibfnamefont{Y.}~\bibnamefont{Tokura}},
  \bibinfo{journal}{Rev. Mod. Phys.} \textbf{\bibinfo{volume}{70}},
  \bibinfo{pages}{1039} (\bibinfo{year}{1998}).

\bibitem[{\citenamefont{Khomskii}(2000)}]{khomskii.pb00}
\bibinfo{author}{\bibfnamefont{D.~I.} \bibnamefont{Khomskii}},
  \bibinfo{journal}{Physica B} \textbf{\bibinfo{volume}{280}},
  \bibinfo{pages}{325} (\bibinfo{year}{2000}).

\bibitem[{\citenamefont{Kagan and Kugel}(2001)}]{kagan.ufn01}
\bibinfo{author}{\bibfnamefont{M.~Y.} \bibnamefont{Kagan}} \bibnamefont{and}
  \bibinfo{author}{\bibfnamefont{K.~I.} \bibnamefont{Kugel}},
  \bibinfo{journal}{Usp. Fiz. Nauk} \textbf{\bibinfo{volume}{171}},
  \bibinfo{pages}{577} (\bibinfo{year}{2001}), \bibinfo{note}{[Phys.\ Usp.\
  {\bf 44}, 553 (2001)]}.

\bibitem[{\citenamefont{Dagotto}(2003)}]{dagotto.03}
\bibinfo{author}{\bibfnamefont{E.}~\bibnamefont{Dagotto}},
  \emph{\bibinfo{title}{Nanoscale Phase Separation and Colossal
  Magnetoresistance: The Physics of Manganites and Related Compounds}}
  (\bibinfo{publisher}{Springer-Verlag}, \bibinfo{address}{Berlin},
  \bibinfo{year}{2003}).

\bibitem[{\citenamefont{Jonker and {\protect Van
  Santen}}(1953)}]{jonker.vansanten.53}
\bibinfo{author}{\bibfnamefont{G.~H.} \bibnamefont{Jonker}} \bibnamefont{and}
  \bibinfo{author}{\bibfnamefont{J.~H.} \bibnamefont{{\protect Van Santen}}},
  \bibinfo{journal}{Physica (Amsterdam)} \textbf{\bibinfo{volume}{19}},
  \bibinfo{pages}{120} (\bibinfo{year}{1953}).

\bibitem[{\citenamefont{Goodenough and Raccah}(1965)}]{goodenough.jap65}
\bibinfo{author}{\bibfnamefont{J.~B.} \bibnamefont{Goodenough}}
  \bibnamefont{and} \bibinfo{author}{\bibfnamefont{P.~M.}
  \bibnamefont{Raccah}}, \bibinfo{journal}{J. Appl. Phys.}
  \textbf{\bibinfo{volume}{36}}, \bibinfo{pages}{1031} (\bibinfo{year}{1965}).

\bibitem[{\citenamefont{Se{\~ n}ar{\' i}s-Rodr{\' i}guez and
  Goodenough}(1995)}]{senarisrodriguez.jssc95}
\bibinfo{author}{\bibfnamefont{M.~A.} \bibnamefont{Se{\~ n}ar{\' i}s-Rodr{\'
  i}guez}} \bibnamefont{and} \bibinfo{author}{\bibfnamefont{J.~B.}
  \bibnamefont{Goodenough}}, \bibinfo{journal}{J. Solid State Chem.}
  \textbf{\bibinfo{volume}{116}}, \bibinfo{pages}{224} (\bibinfo{year}{1995}).

\bibitem[{\citenamefont{Asai et~al.}(1998)\citenamefont{Asai, Yoneda, Yokokura,
  Tranquada, Shirane, and Kohn}}]{asai.jpsj98}
\bibinfo{author}{\bibfnamefont{K.}~\bibnamefont{Asai}},
  \bibinfo{author}{\bibfnamefont{A.}~\bibnamefont{Yoneda}},
  \bibinfo{author}{\bibfnamefont{O.}~\bibnamefont{Yokokura}},
  \bibinfo{author}{\bibfnamefont{J.~M.} \bibnamefont{Tranquada}},
  \bibinfo{author}{\bibfnamefont{G.}~\bibnamefont{Shirane}}, \bibnamefont{and}
  \bibinfo{author}{\bibfnamefont{K.}~\bibnamefont{Kohn}}, \bibinfo{journal}{J.
  Phys. Soc. Jpn.} \textbf{\bibinfo{volume}{67}}, \bibinfo{pages}{290}
  (\bibinfo{year}{1998}).

\bibitem[{\citenamefont{Saitoh et~al.}(1997)\citenamefont{Saitoh, Mizokawa,
  Fujimori, Abbate, Takeda, and Takano}}]{saitoh.prb97}
\bibinfo{author}{\bibfnamefont{T.}~\bibnamefont{Saitoh}},
  \bibinfo{author}{\bibfnamefont{T.}~\bibnamefont{Mizokawa}},
  \bibinfo{author}{\bibfnamefont{A.}~\bibnamefont{Fujimori}},
  \bibinfo{author}{\bibfnamefont{M.}~\bibnamefont{Abbate}},
  \bibinfo{author}{\bibfnamefont{Y.}~\bibnamefont{Takeda}}, \bibnamefont{and}
  \bibinfo{author}{\bibfnamefont{M.}~\bibnamefont{Takano}},
  \bibinfo{journal}{Phys. Rev. B} \textbf{\bibinfo{volume}{55}},
  \bibinfo{pages}{4257} (\bibinfo{year}{1997}).

\bibitem[{\citenamefont{Tokura et~al.}(1998)\citenamefont{Tokura, Okimoto,
  Yamaguchi, Taniguchi, Kimura, and Takagi}}]{tokura.prb98}
\bibinfo{author}{\bibfnamefont{Y.}~\bibnamefont{Tokura}},
  \bibinfo{author}{\bibfnamefont{Y.}~\bibnamefont{Okimoto}},
  \bibinfo{author}{\bibfnamefont{S.}~\bibnamefont{Yamaguchi}},
  \bibinfo{author}{\bibfnamefont{H.}~\bibnamefont{Taniguchi}},
  \bibinfo{author}{\bibfnamefont{T.}~\bibnamefont{Kimura}}, \bibnamefont{and}
  \bibinfo{author}{\bibfnamefont{H.}~\bibnamefont{Takagi}},
  \bibinfo{journal}{Phys. Rev. B} \textbf{\bibinfo{volume}{58}},
  \bibinfo{pages}{R1699} (\bibinfo{year}{1998}).

\bibitem[{\citenamefont{Korotin et~al.}(1996)\citenamefont{Korotin, Ezhov,
  Solovyev, Anisimov, Khomskii, and Sawatzky}}]{korotin.prb96}
\bibinfo{author}{\bibfnamefont{M.~A.} \bibnamefont{Korotin}},
  \bibinfo{author}{\bibfnamefont{S.~Y.} \bibnamefont{Ezhov}},
  \bibinfo{author}{\bibfnamefont{I.~V.} \bibnamefont{Solovyev}},
  \bibinfo{author}{\bibfnamefont{V.~I.} \bibnamefont{Anisimov}},
  \bibinfo{author}{\bibfnamefont{D.~I.} \bibnamefont{Khomskii}},
  \bibnamefont{and} \bibinfo{author}{\bibfnamefont{G.~A.}
  \bibnamefont{Sawatzky}}, \bibinfo{journal}{Phys. Rev. B}
  \textbf{\bibinfo{volume}{54}}, \bibinfo{pages}{5309} (\bibinfo{year}{1996}).

\bibitem[{\citenamefont{Yamaguchi et~al.}(1997)\citenamefont{Yamaguchi,
  Okimoto, and Tokura}}]{yamaguchi.prb97}
\bibinfo{author}{\bibfnamefont{S.}~\bibnamefont{Yamaguchi}},
  \bibinfo{author}{\bibfnamefont{Y.}~\bibnamefont{Okimoto}}, \bibnamefont{and}
  \bibinfo{author}{\bibfnamefont{Y.}~\bibnamefont{Tokura}},
  \bibinfo{journal}{Phys. Rev. B} \textbf{\bibinfo{volume}{55}},
  \bibinfo{pages}{R8666} (\bibinfo{year}{1997}).

\bibitem[{\citenamefont{Kobayashi et~al.}(2000)\citenamefont{Kobayashi,
  Fujiwara, Murata, Asai, and Yasuoka}}]{kobayashi.prb00}
\bibinfo{author}{\bibfnamefont{Y.}~\bibnamefont{Kobayashi}},
  \bibinfo{author}{\bibfnamefont{N.}~\bibnamefont{Fujiwara}},
  \bibinfo{author}{\bibfnamefont{S.}~\bibnamefont{Murata}},
  \bibinfo{author}{\bibfnamefont{K.}~\bibnamefont{Asai}}, \bibnamefont{and}
  \bibinfo{author}{\bibfnamefont{H.}~\bibnamefont{Yasuoka}},
  \bibinfo{journal}{Phys. Rev. B} \textbf{\bibinfo{volume}{62}},
  \bibinfo{pages}{410} (\bibinfo{year}{2000}).

\bibitem[{\citenamefont{Sato et~al.}(2008)\citenamefont{Sato, Bartashevich,
  Goto, Kobayashi, Suzuki, Asai, Matsuo, and Kindo}}]{sato.jpsj08}
\bibinfo{author}{\bibfnamefont{K.}~\bibnamefont{Sato}},
  \bibinfo{author}{\bibfnamefont{M.~I.} \bibnamefont{Bartashevich}},
  \bibinfo{author}{\bibfnamefont{T.}~\bibnamefont{Goto}},
  \bibinfo{author}{\bibfnamefont{Y.}~\bibnamefont{Kobayashi}},
  \bibinfo{author}{\bibfnamefont{M.}~\bibnamefont{Suzuki}},
  \bibinfo{author}{\bibfnamefont{K.}~\bibnamefont{Asai}},
  \bibinfo{author}{\bibfnamefont{A.}~\bibnamefont{Matsuo}}, \bibnamefont{and}
  \bibinfo{author}{\bibfnamefont{K.}~\bibnamefont{Kindo}}, \bibinfo{journal}{J.
  Phys. Soc. Jpn.} \textbf{\bibinfo{volume}{77}}, \bibinfo{pages}{024601}
  (\bibinfo{year}{2008}).

\bibitem[{\citenamefont{Zobel et~al.}(2002)\citenamefont{Zobel, Kriener, Bruns,
  Baier, Gr\"uninger, Lorenz, Reutler, and Revcolevschi}}]{zobel.prb02}
\bibinfo{author}{\bibfnamefont{C.}~\bibnamefont{Zobel}},
  \bibinfo{author}{\bibfnamefont{M.}~\bibnamefont{Kriener}},
  \bibinfo{author}{\bibfnamefont{D.}~\bibnamefont{Bruns}},
  \bibinfo{author}{\bibfnamefont{J.}~\bibnamefont{Baier}},
  \bibinfo{author}{\bibfnamefont{M.}~\bibnamefont{Gr\"uninger}},
  \bibinfo{author}{\bibfnamefont{T.}~\bibnamefont{Lorenz}},
  \bibinfo{author}{\bibfnamefont{P.}~\bibnamefont{Reutler}}, \bibnamefont{and}
  \bibinfo{author}{\bibfnamefont{A.}~\bibnamefont{Revcolevschi}},
  \bibinfo{journal}{Phys. Rev. B} \textbf{\bibinfo{volume}{66}},
  \bibinfo{pages}{020402(R)} (\bibinfo{year}{2002}).

\bibitem[{\citenamefont{Berggold et~al.}(2008)\citenamefont{Berggold, Kriener,
  Becker, Benomar, Reuther, Zobel, and Lorenz}}]{berggold.prb08}
\bibinfo{author}{\bibfnamefont{K.}~\bibnamefont{Berggold}},
  \bibinfo{author}{\bibfnamefont{M.}~\bibnamefont{Kriener}},
  \bibinfo{author}{\bibfnamefont{P.}~\bibnamefont{Becker}},
  \bibinfo{author}{\bibfnamefont{M.}~\bibnamefont{Benomar}},
  \bibinfo{author}{\bibfnamefont{M.}~\bibnamefont{Reuther}},
  \bibinfo{author}{\bibfnamefont{C.}~\bibnamefont{Zobel}}, \bibnamefont{and}
  \bibinfo{author}{\bibfnamefont{T.}~\bibnamefont{Lorenz}},
  \bibinfo{journal}{Phys. Rev. B} \textbf{\bibinfo{volume}{78}},
  \bibinfo{eid}{134402} (\bibinfo{year}{2008}).

\bibitem[{\citenamefont{Baier et~al.}(2005)\citenamefont{Baier, Jodlauk,
  Kriener, Reichl, Zobel, Kierspel, Freimuth, and Lorenz}}]{baier.prb05}
\bibinfo{author}{\bibfnamefont{J.}~\bibnamefont{Baier}},
  \bibinfo{author}{\bibfnamefont{S.}~\bibnamefont{Jodlauk}},
  \bibinfo{author}{\bibfnamefont{M.}~\bibnamefont{Kriener}},
  \bibinfo{author}{\bibfnamefont{A.}~\bibnamefont{Reichl}},
  \bibinfo{author}{\bibfnamefont{C.}~\bibnamefont{Zobel}},
  \bibinfo{author}{\bibfnamefont{H.}~\bibnamefont{Kierspel}},
  \bibinfo{author}{\bibfnamefont{A.}~\bibnamefont{Freimuth}}, \bibnamefont{and}
  \bibinfo{author}{\bibfnamefont{T.}~\bibnamefont{Lorenz}},
  \bibinfo{journal}{Phys. Rev. B} \textbf{\bibinfo{volume}{71}},
  \bibinfo{pages}{014443} (\bibinfo{year}{2005}).

\bibitem[{\citenamefont{Tong et~al.}(2009)\citenamefont{Tong, Wu, Kim, Kwon,
  Park, and Kim}}]{tong.jpsj09}
\bibinfo{author}{\bibfnamefont{P.}~\bibnamefont{Tong}},
  \bibinfo{author}{\bibfnamefont{Y.}~\bibnamefont{Wu}},
  \bibinfo{author}{\bibfnamefont{B.}~\bibnamefont{Kim}},
  \bibinfo{author}{\bibfnamefont{D.}~\bibnamefont{Kwon}},
  \bibinfo{author}{\bibfnamefont{J.~M.~S.} \bibnamefont{Park}},
  \bibnamefont{and} \bibinfo{author}{\bibfnamefont{B.~G.} \bibnamefont{Kim}},
  \bibinfo{journal}{J. Phys. Soc. Jpn.} \textbf{\bibinfo{volume}{78}},
  \bibinfo{pages}{034702} (\bibinfo{year}{2009}).

\bibitem[{\citenamefont{Tsubouchi et~al.}(2002)\citenamefont{Tsubouchi,
  Ky{\^o}men, Itoh, Ganguly, Oguni, Shimojo, Morii, and
  Ishii}}]{tsubouchi.prb02}
\bibinfo{author}{\bibfnamefont{S.}~\bibnamefont{Tsubouchi}},
  \bibinfo{author}{\bibfnamefont{T.}~\bibnamefont{Ky{\^o}men}},
  \bibinfo{author}{\bibfnamefont{M.}~\bibnamefont{Itoh}},
  \bibinfo{author}{\bibfnamefont{P.}~\bibnamefont{Ganguly}},
  \bibinfo{author}{\bibfnamefont{M.}~\bibnamefont{Oguni}},
  \bibinfo{author}{\bibfnamefont{Y.}~\bibnamefont{Shimojo}},
  \bibinfo{author}{\bibfnamefont{Y.}~\bibnamefont{Morii}}, \bibnamefont{and}
  \bibinfo{author}{\bibfnamefont{Y.}~\bibnamefont{Ishii}},
  \bibinfo{journal}{Phys. Rev. B} \textbf{\bibinfo{volume}{66}},
  \bibinfo{pages}{052418} (\bibinfo{year}{2002}).

\bibitem[{\citenamefont{Fujita et~al.}(2004)\citenamefont{Fujita, Miyashita,
  Yasui, Kobayashi, Sato, Nishibori, Sakata, Shimojo, Igawa, Ishii
  et~al.}}]{fujita.jpsj04}
\bibinfo{author}{\bibfnamefont{T.}~\bibnamefont{Fujita}},
  \bibinfo{author}{\bibfnamefont{T.}~\bibnamefont{Miyashita}},
  \bibinfo{author}{\bibfnamefont{Y.}~\bibnamefont{Yasui}},
  \bibinfo{author}{\bibfnamefont{Y.}~\bibnamefont{Kobayashi}},
  \bibinfo{author}{\bibfnamefont{M.}~\bibnamefont{Sato}},
  \bibinfo{author}{\bibfnamefont{E.}~\bibnamefont{Nishibori}},
  \bibinfo{author}{\bibfnamefont{M.}~\bibnamefont{Sakata}},
  \bibinfo{author}{\bibfnamefont{Y.}~\bibnamefont{Shimojo}},
  \bibinfo{author}{\bibfnamefont{N.}~\bibnamefont{Igawa}},
  \bibinfo{author}{\bibfnamefont{Y.}~\bibnamefont{Ishii}},
  \bibnamefont{et~al.}, \bibinfo{journal}{J. Phys. Soc. Jpn.}
  \textbf{\bibinfo{volume}{73}}, \bibinfo{pages}{1987} (\bibinfo{year}{2004}).

\bibitem[{\citenamefont{Podlesnyak et~al.}(2006)\citenamefont{Podlesnyak,
  Streule, Mesot, Medarde, Pomjakushina, Conder, Tanaka, Haverkort, and
  Khomskii}}]{podlesnyak.prl06}
\bibinfo{author}{\bibfnamefont{A.}~\bibnamefont{Podlesnyak}},
  \bibinfo{author}{\bibfnamefont{S.}~\bibnamefont{Streule}},
  \bibinfo{author}{\bibfnamefont{J.}~\bibnamefont{Mesot}},
  \bibinfo{author}{\bibfnamefont{M.}~\bibnamefont{Medarde}},
  \bibinfo{author}{\bibfnamefont{E.}~\bibnamefont{Pomjakushina}},
  \bibinfo{author}{\bibfnamefont{K.}~\bibnamefont{Conder}},
  \bibinfo{author}{\bibfnamefont{A.}~\bibnamefont{Tanaka}},
  \bibinfo{author}{\bibfnamefont{M.~W.} \bibnamefont{Haverkort}},
  \bibnamefont{and} \bibinfo{author}{\bibfnamefont{D.~I.}
  \bibnamefont{Khomskii}}, \bibinfo{journal}{Phys. Rev. Lett.}
  \textbf{\bibinfo{volume}{97}}, \bibinfo{eid}{247208} (\bibinfo{year}{2006}).

\bibitem[{\citenamefont{Haverkort et~al.}(2006)\citenamefont{Haverkort, Hu,
  Cezar, Burnus, Hartmann, Reuther, Zobel, Lorenz, Tanaka, Brookes
  et~al.}}]{haverkort.prl06}
\bibinfo{author}{\bibfnamefont{M.~W.} \bibnamefont{Haverkort}},
  \bibinfo{author}{\bibfnamefont{Z.}~\bibnamefont{Hu}},
  \bibinfo{author}{\bibfnamefont{J.~C.} \bibnamefont{Cezar}},
  \bibinfo{author}{\bibfnamefont{T.}~\bibnamefont{Burnus}},
  \bibinfo{author}{\bibfnamefont{H.}~\bibnamefont{Hartmann}},
  \bibinfo{author}{\bibfnamefont{M.}~\bibnamefont{Reuther}},
  \bibinfo{author}{\bibfnamefont{C.}~\bibnamefont{Zobel}},
  \bibinfo{author}{\bibfnamefont{T.}~\bibnamefont{Lorenz}},
  \bibinfo{author}{\bibfnamefont{A.}~\bibnamefont{Tanaka}},
  \bibinfo{author}{\bibfnamefont{N.~B.} \bibnamefont{Brookes}},
  \bibnamefont{et~al.}, \bibinfo{journal}{Phys. Rev. Lett.}
  \textbf{\bibinfo{volume}{97}}, \bibinfo{eid}{176405} (\bibinfo{year}{2006}).

\bibitem[{\citenamefont{Ropka and Radwanski}(2003)}]{ropka.prb03}
\bibinfo{author}{\bibfnamefont{Z.}~\bibnamefont{Ropka}} \bibnamefont{and}
  \bibinfo{author}{\bibfnamefont{R.~J.} \bibnamefont{Radwanski}},
  \bibinfo{journal}{Phys. Rev. B} \textbf{\bibinfo{volume}{67}},
  \bibinfo{pages}{172401} (\bibinfo{year}{2003}).

\bibitem[{\citenamefont{Haverkort}(2005)}]{haverkort.phd}
\bibinfo{author}{\bibfnamefont{M.~W.} \bibnamefont{Haverkort}}, Ph.D. thesis,
  \bibinfo{school}{University of Cologne} (\bibinfo{year}{2005}),
  \bibinfo{note}{cond-mat/0505214}.

\bibitem[{\citenamefont{Noguchi et~al.}(2002)\citenamefont{Noguchi, Kawamata,
  Okuda, Nojiri, and Motokawa}}]{noguchi.prb02}
\bibinfo{author}{\bibfnamefont{S.}~\bibnamefont{Noguchi}},
  \bibinfo{author}{\bibfnamefont{S.}~\bibnamefont{Kawamata}},
  \bibinfo{author}{\bibfnamefont{K.}~\bibnamefont{Okuda}},
  \bibinfo{author}{\bibfnamefont{H.}~\bibnamefont{Nojiri}}, \bibnamefont{and}
  \bibinfo{author}{\bibfnamefont{M.}~\bibnamefont{Motokawa}},
  \bibinfo{journal}{Phys. Rev. B} \textbf{\bibinfo{volume}{66}},
  \bibinfo{pages}{094404} (\bibinfo{year}{2002}).

\bibitem[{\citenamefont{Phelan et~al.}(2008)\citenamefont{Phelan, Yu, and
  Louca}}]{phelan.prb08}
\bibinfo{author}{\bibfnamefont{D.}~\bibnamefont{Phelan}},
  \bibinfo{author}{\bibfnamefont{J.}~\bibnamefont{Yu}}, \bibnamefont{and}
  \bibinfo{author}{\bibfnamefont{D.}~\bibnamefont{Louca}},
  \bibinfo{journal}{Phys. Rev. B} \textbf{\bibinfo{volume}{78}},
  \bibinfo{eid}{094108} (\bibinfo{year}{2008}).

\bibitem[{\citenamefont{Phelan et~al.}(2009)\citenamefont{Phelan, Louca,
  Ancona, Rosenkranz, Zheng, and Mitchell}}]{phelan.prb09}
\bibinfo{author}{\bibfnamefont{D.}~\bibnamefont{Phelan}},
  \bibinfo{author}{\bibfnamefont{D.}~\bibnamefont{Louca}},
  \bibinfo{author}{\bibfnamefont{S.~N.} \bibnamefont{Ancona}},
  \bibinfo{author}{\bibfnamefont{S.}~\bibnamefont{Rosenkranz}},
  \bibinfo{author}{\bibfnamefont{H.}~\bibnamefont{Zheng}}, \bibnamefont{and}
  \bibinfo{author}{\bibfnamefont{J.~F.} \bibnamefont{Mitchell}},
  \bibinfo{journal}{Phys. Rev. B} \textbf{\bibinfo{volume}{79}},
  \bibinfo{eid}{094420} (\bibinfo{year}{2009}).

\bibitem[{\citenamefont{Podlesnyak et~al.}(2008)\citenamefont{Podlesnyak,
  Russina, Furrer, Alfonsov, Vavilova, Kataev, B\"{u}chner, Str\"{a}ssle,
  Pomjakushina, Conder et~al.}}]{podlesnyak.prl08}
\bibinfo{author}{\bibfnamefont{A.}~\bibnamefont{Podlesnyak}},
  \bibinfo{author}{\bibfnamefont{M.}~\bibnamefont{Russina}},
  \bibinfo{author}{\bibfnamefont{A.}~\bibnamefont{Furrer}},
  \bibinfo{author}{\bibfnamefont{A.}~\bibnamefont{Alfonsov}},
  \bibinfo{author}{\bibfnamefont{E.}~\bibnamefont{Vavilova}},
  \bibinfo{author}{\bibfnamefont{V.}~\bibnamefont{Kataev}},
  \bibinfo{author}{\bibfnamefont{B.}~\bibnamefont{B\"{u}chner}},
  \bibinfo{author}{\bibfnamefont{T.}~\bibnamefont{Str\"{a}ssle}},
  \bibinfo{author}{\bibfnamefont{E.}~\bibnamefont{Pomjakushina}},
  \bibinfo{author}{\bibfnamefont{K.}~\bibnamefont{Conder}},
  \bibnamefont{et~al.}, \bibinfo{journal}{Phys. Rev. Lett.}
  \textbf{\bibinfo{volume}{101}}, \bibinfo{eid}{247603} (\bibinfo{year}{2008}).

\bibitem[{\citenamefont{Maignan et~al.}(2004)\citenamefont{Maignan, Caignaert,
  Raveau, Khomskii, and Sawatzky}}]{maignan.prl04}
\bibinfo{author}{\bibfnamefont{A.}~\bibnamefont{Maignan}},
  \bibinfo{author}{\bibfnamefont{V.}~\bibnamefont{Caignaert}},
  \bibinfo{author}{\bibfnamefont{B.}~\bibnamefont{Raveau}},
  \bibinfo{author}{\bibfnamefont{D.}~\bibnamefont{Khomskii}}, \bibnamefont{and}
  \bibinfo{author}{\bibfnamefont{G.}~\bibnamefont{Sawatzky}},
  \bibinfo{journal}{Phys. Rev. Lett.} \textbf{\bibinfo{volume}{93}},
  \bibinfo{pages}{026401} (\bibinfo{year}{2004}).

\bibitem[{\citenamefont{Kriener et~al.}(2004)\citenamefont{Kriener, Zobel,
  Reichl, Baier, Cwik, Berggold, Kierspel, Zabara, Freimuth, and
  Lorenz}}]{kriener.prb04}
\bibinfo{author}{\bibfnamefont{M.}~\bibnamefont{Kriener}},
  \bibinfo{author}{\bibfnamefont{C.}~\bibnamefont{Zobel}},
  \bibinfo{author}{\bibfnamefont{A.}~\bibnamefont{Reichl}},
  \bibinfo{author}{\bibfnamefont{J.}~\bibnamefont{Baier}},
  \bibinfo{author}{\bibfnamefont{M.}~\bibnamefont{Cwik}},
  \bibinfo{author}{\bibfnamefont{K.}~\bibnamefont{Berggold}},
  \bibinfo{author}{\bibfnamefont{H.}~\bibnamefont{Kierspel}},
  \bibinfo{author}{\bibfnamefont{O.}~\bibnamefont{Zabara}},
  \bibinfo{author}{\bibfnamefont{A.}~\bibnamefont{Freimuth}}, \bibnamefont{and}
  \bibinfo{author}{\bibfnamefont{T.}~\bibnamefont{Lorenz}},
  \bibinfo{journal}{Phys. Rev. B} \textbf{\bibinfo{volume}{69}},
  \bibinfo{pages}{094417} (\bibinfo{year}{2004}).

\bibitem[{\citenamefont{Zener}(1951)}]{zener.pr51}
\bibinfo{author}{\bibfnamefont{C.}~\bibnamefont{Zener}},
  \bibinfo{journal}{Phys. Rev.} \textbf{\bibinfo{volume}{82}},
  \bibinfo{pages}{403} (\bibinfo{year}{1951}).

\bibitem[{\citenamefont{de~Gennes}(1960)}]{degennes.pr60}
\bibinfo{author}{\bibfnamefont{P.~G.} \bibnamefont{de~Gennes}},
  \bibinfo{journal}{Phys. Rev.} \textbf{\bibinfo{volume}{118}},
  \bibinfo{pages}{141} (\bibinfo{year}{1960}).

\bibitem[{\citenamefont{Kriener et~al.}(2009)\citenamefont{Kriener, Braden,
  Kierspel, Senff, Zabara, Zobel, and Lorenz}}]{kriener.prb09}
\bibinfo{author}{\bibfnamefont{M.}~\bibnamefont{Kriener}},
  \bibinfo{author}{\bibfnamefont{M.}~\bibnamefont{Braden}},
  \bibinfo{author}{\bibfnamefont{H.}~\bibnamefont{Kierspel}},
  \bibinfo{author}{\bibfnamefont{D.}~\bibnamefont{Senff}},
  \bibinfo{author}{\bibfnamefont{O.}~\bibnamefont{Zabara}},
  \bibinfo{author}{\bibfnamefont{C.}~\bibnamefont{Zobel}}, \bibnamefont{and}
  \bibinfo{author}{\bibfnamefont{T.}~\bibnamefont{Lorenz}},
  \bibinfo{journal}{Phys. Rev. B} \textbf{\bibinfo{volume}{79}},
  \bibinfo{eid}{224104} (\bibinfo{year}{2009}).

\bibitem[{\citenamefont{Fujita et~al.}(2005)\citenamefont{Fujita, Kawabata,
  Sato, Kurita, Hedo, and Uwatoko}}]{fujita.jpsj05}
\bibinfo{author}{\bibfnamefont{T.}~\bibnamefont{Fujita}},
  \bibinfo{author}{\bibfnamefont{S.}~\bibnamefont{Kawabata}},
  \bibinfo{author}{\bibfnamefont{M.}~\bibnamefont{Sato}},
  \bibinfo{author}{\bibfnamefont{N.}~\bibnamefont{Kurita}},
  \bibinfo{author}{\bibfnamefont{M.}~\bibnamefont{Hedo}}, \bibnamefont{and}
  \bibinfo{author}{\bibfnamefont{Y.}~\bibnamefont{Uwatoko}},
  \bibinfo{journal}{J. Phys. Soc. Jpn.} \textbf{\bibinfo{volume}{74}},
  \bibinfo{pages}{2294} (\bibinfo{year}{2005}).

\bibitem[{\citenamefont{Babushkina
  et~al.}(1998{\natexlab{a}})\citenamefont{Babushkina, Belova, Gorbenko, Kaul,
  Bosak, Ozhogin, and Kugel}}]{babushkina.nat98}
\bibinfo{author}{\bibfnamefont{N.~A.} \bibnamefont{Babushkina}},
  \bibinfo{author}{\bibfnamefont{L.~M.} \bibnamefont{Belova}},
  \bibinfo{author}{\bibfnamefont{O.~Y.} \bibnamefont{Gorbenko}},
  \bibinfo{author}{\bibfnamefont{A.~R.} \bibnamefont{Kaul}},
  \bibinfo{author}{\bibfnamefont{A.~A.} \bibnamefont{Bosak}},
  \bibinfo{author}{\bibfnamefont{V.~I.} \bibnamefont{Ozhogin}},
  \bibnamefont{and} \bibinfo{author}{\bibfnamefont{K.~I.} \bibnamefont{Kugel}},
  \bibinfo{journal}{Nature (London)} \textbf{\bibinfo{volume}{391}},
  \bibinfo{pages}{159} (\bibinfo{year}{1998}{\natexlab{a}}).

\bibitem[{\citenamefont{Wang et~al.}(2006{\natexlab{a}})\citenamefont{Wang, Wu,
  Luo, Wang, and Chen}}]{wang.prb06}
\bibinfo{author}{\bibfnamefont{G.~Y.} \bibnamefont{Wang}},
  \bibinfo{author}{\bibfnamefont{T.}~\bibnamefont{Wu}},
  \bibinfo{author}{\bibfnamefont{X.~G.} \bibnamefont{Luo}},
  \bibinfo{author}{\bibfnamefont{W.}~\bibnamefont{Wang}}, \bibnamefont{and}
  \bibinfo{author}{\bibfnamefont{X.~H.} \bibnamefont{Chen}},
  \bibinfo{journal}{Phys. Rev. B} \textbf{\bibinfo{volume}{73}},
  \bibinfo{eid}{052404} (\bibinfo{year}{2006}{\natexlab{a}}).

\bibitem[{\citenamefont{Wang et~al.}(2006{\natexlab{b}})\citenamefont{Wang,
  Chen, Wu, Wu, Luo, and Wang}}]{wang.prb06.2}
\bibinfo{author}{\bibfnamefont{G.~Y.} \bibnamefont{Wang}},
  \bibinfo{author}{\bibfnamefont{X.~H.} \bibnamefont{Chen}},
  \bibinfo{author}{\bibfnamefont{T.}~\bibnamefont{Wu}},
  \bibinfo{author}{\bibfnamefont{G.}~\bibnamefont{Wu}},
  \bibinfo{author}{\bibfnamefont{X.~G.} \bibnamefont{Luo}}, \bibnamefont{and}
  \bibinfo{author}{\bibfnamefont{C.~H.} \bibnamefont{Wang}},
  \bibinfo{journal}{Phys. Rev. B} \textbf{\bibinfo{volume}{74}},
  \bibinfo{eid}{165113} (\bibinfo{year}{2006}{\natexlab{b}}).

\bibitem[{\citenamefont{Shannon}(1976)}]{shannon.76}
\bibinfo{author}{\bibfnamefont{R.~D.} \bibnamefont{Shannon}},
  \bibinfo{journal}{Acta Cryst. A} \textbf{\bibinfo{volume}{32}},
  \bibinfo{pages}{751} (\bibinfo{year}{1976}).

\bibitem[{\citenamefont{Balagurov et~al.}(1999)\citenamefont{Balagurov,
  Pomjakushin, Sheptyakov, Aksenov, Babushkina, Belova, Taldenkov, Inyushkin,
  Fischer, Gutmann et~al.}}]{balagurov.prb99}
\bibinfo{author}{\bibfnamefont{A.~M.} \bibnamefont{Balagurov}},
  \bibinfo{author}{\bibfnamefont{V.~Y.} \bibnamefont{Pomjakushin}},
  \bibinfo{author}{\bibfnamefont{D.~V.} \bibnamefont{Sheptyakov}},
  \bibinfo{author}{\bibfnamefont{V.~L.} \bibnamefont{Aksenov}},
  \bibinfo{author}{\bibfnamefont{N.~A.} \bibnamefont{Babushkina}},
  \bibinfo{author}{\bibfnamefont{L.~M.} \bibnamefont{Belova}},
  \bibinfo{author}{\bibfnamefont{A.~N.} \bibnamefont{Taldenkov}},
  \bibinfo{author}{\bibfnamefont{A.~V.} \bibnamefont{Inyushkin}},
  \bibinfo{author}{\bibfnamefont{P.}~\bibnamefont{Fischer}},
  \bibinfo{author}{\bibfnamefont{M.}~\bibnamefont{Gutmann}},
  \bibnamefont{et~al.}, \bibinfo{journal}{Phys. Rev. B}
  \textbf{\bibinfo{volume}{60}}, \bibinfo{pages}{383} (\bibinfo{year}{1999}).

\bibitem[{\citenamefont{Babushkina
  et~al.}(1998{\natexlab{b}})\citenamefont{Babushkina, Belova, Ozhogin,
  Gorbenko, Kaul, Bosak, Khomskii, and Kugel}}]{babushkina.jap98}
\bibinfo{author}{\bibfnamefont{N.~A.} \bibnamefont{Babushkina}},
  \bibinfo{author}{\bibfnamefont{L.~M.} \bibnamefont{Belova}},
  \bibinfo{author}{\bibfnamefont{V.~I.} \bibnamefont{Ozhogin}},
  \bibinfo{author}{\bibfnamefont{O.~Y.} \bibnamefont{Gorbenko}},
  \bibinfo{author}{\bibfnamefont{A.~R.} \bibnamefont{Kaul}},
  \bibinfo{author}{\bibfnamefont{A.~A.} \bibnamefont{Bosak}},
  \bibinfo{author}{\bibfnamefont{D.~I.} \bibnamefont{Khomskii}},
  \bibnamefont{and} \bibinfo{author}{\bibfnamefont{K.~I.} \bibnamefont{Kugel}},
  \bibinfo{journal}{J. Appl. Phys.} \textbf{\bibinfo{volume}{83}},
  \bibinfo{pages}{7369} (\bibinfo{year}{1998}{\natexlab{b}}).

\bibitem[{\citenamefont{Brinks et~al.}(1999)\citenamefont{Brinks, Fjellv{\aa}g,
  Kjekshus, and Hauback}}]{brinks.jssch99}
\bibinfo{author}{\bibfnamefont{H.~W.} \bibnamefont{Brinks}},
  \bibinfo{author}{\bibfnamefont{H.}~\bibnamefont{Fjellv{\aa}g}},
  \bibinfo{author}{\bibfnamefont{A.}~\bibnamefont{Kjekshus}}, \bibnamefont{and}
  \bibinfo{author}{\bibfnamefont{B.~C.} \bibnamefont{Hauback}},
  \bibinfo{journal}{J. Solid State Chem.} \textbf{\bibinfo{volume}{147}},
  \bibinfo{pages}{464} (\bibinfo{year}{1999}).

\bibitem[{\citenamefont{Mineshige et~al.}(1996)\citenamefont{Mineshige, Inaba,
  Yao, Ogumi, Kikuchi, and Kawase}}]{mineshige.jssch96}
\bibinfo{author}{\bibfnamefont{A.}~\bibnamefont{Mineshige}},
  \bibinfo{author}{\bibfnamefont{M.}~\bibnamefont{Inaba}},
  \bibinfo{author}{\bibfnamefont{T.}~\bibnamefont{Yao}},
  \bibinfo{author}{\bibfnamefont{Z.}~\bibnamefont{Ogumi}},
  \bibinfo{author}{\bibfnamefont{K.}~\bibnamefont{Kikuchi}}, \bibnamefont{and}
  \bibinfo{author}{\bibfnamefont{M.}~\bibnamefont{Kawase}},
  \bibinfo{journal}{J. Solid State Chem.} \textbf{\bibinfo{volume}{121}},
  \bibinfo{pages}{423} (\bibinfo{year}{1996}).

\bibitem[{\citenamefont{Radaelli and Cheong}(2002)}]{radaelli.prb02}
\bibinfo{author}{\bibfnamefont{P.~G.} \bibnamefont{Radaelli}} \bibnamefont{and}
  \bibinfo{author}{\bibfnamefont{S.-W.} \bibnamefont{Cheong}},
  \bibinfo{journal}{Phys. Rev. B} \textbf{\bibinfo{volume}{66}},
  \bibinfo{pages}{094408} (\bibinfo{year}{2002}).

\bibitem[{\citenamefont{Serrano et~al.}(2006)\citenamefont{Serrano, Kremer,
  Cardona, Siegle, Romero, and Lauck}}]{serrano.prb06.ZnO}
\bibinfo{author}{\bibfnamefont{J.}~\bibnamefont{Serrano}},
  \bibinfo{author}{\bibfnamefont{R.~K.} \bibnamefont{Kremer}},
  \bibinfo{author}{\bibfnamefont{M.}~\bibnamefont{Cardona}},
  \bibinfo{author}{\bibfnamefont{G.}~\bibnamefont{Siegle}},
  \bibinfo{author}{\bibfnamefont{A.~H.} \bibnamefont{Romero}},
  \bibnamefont{and} \bibinfo{author}{\bibfnamefont{R.}~\bibnamefont{Lauck}},
  \bibinfo{journal}{Phys. Rev. B} \textbf{\bibinfo{volume}{73}},
  \bibinfo{eid}{094303} (\bibinfo{year}{2006}).

\bibitem[{\citenamefont{Cardona et~al.}(2007)\citenamefont{Cardona, Kremer,
  Lauck, Siegle, Serrano, and Romero}}]{cardona.prb07.PbS}
\bibinfo{author}{\bibfnamefont{M.}~\bibnamefont{Cardona}},
  \bibinfo{author}{\bibfnamefont{R.~K.} \bibnamefont{Kremer}},
  \bibinfo{author}{\bibfnamefont{R.}~\bibnamefont{Lauck}},
  \bibinfo{author}{\bibfnamefont{G.}~\bibnamefont{Siegle}},
  \bibinfo{author}{\bibfnamefont{J.}~\bibnamefont{Serrano}}, \bibnamefont{and}
  \bibinfo{author}{\bibfnamefont{A.~H.} \bibnamefont{Romero}},
  \bibinfo{journal}{Phys. Rev. B} \textbf{\bibinfo{volume}{76}},
  \bibinfo{eid}{075211} (\bibinfo{year}{2007}).

\bibitem[{\citenamefont{Paraskevopoulos
  et~al.}(2001)\citenamefont{Paraskevopoulos, Hemberger, Krimmel, and
  Loidl}}]{paraskevopoulos.prb01}
\bibinfo{author}{\bibfnamefont{M.}~\bibnamefont{Paraskevopoulos}},
  \bibinfo{author}{\bibfnamefont{J.}~\bibnamefont{Hemberger}},
  \bibinfo{author}{\bibfnamefont{A.}~\bibnamefont{Krimmel}}, \bibnamefont{and}
  \bibinfo{author}{\bibfnamefont{A.}~\bibnamefont{Loidl}},
  \bibinfo{journal}{Phys. Rev. B} \textbf{\bibinfo{volume}{63}},
  \bibinfo{pages}{224416} (\bibinfo{year}{2001}).

\bibitem[{\citenamefont{Aarbogh et~al.}(2006)\citenamefont{Aarbogh, Wu, Wang,
  Zheng, Mitchell, and Leighton}}]{aarbogh.prb06}
\bibinfo{author}{\bibfnamefont{H.~M.} \bibnamefont{Aarbogh}},
  \bibinfo{author}{\bibfnamefont{J.}~\bibnamefont{Wu}},
  \bibinfo{author}{\bibfnamefont{L.}~\bibnamefont{Wang}},
  \bibinfo{author}{\bibfnamefont{H.}~\bibnamefont{Zheng}},
  \bibinfo{author}{\bibfnamefont{J.~F.} \bibnamefont{Mitchell}},
  \bibnamefont{and} \bibinfo{author}{\bibfnamefont{C.}~\bibnamefont{Leighton}},
  \bibinfo{journal}{Phys. Rev. B} \textbf{\bibinfo{volume}{74}},
  \bibinfo{eid}{134408} (\bibinfo{year}{2006}).

\bibitem[{\citenamefont{Tsubouchi et~al.}(2004)\citenamefont{Tsubouchi,
  Ky\^omen, Itoh, and Oguni}}]{tsubouchi.prb04}
\bibinfo{author}{\bibfnamefont{S.}~\bibnamefont{Tsubouchi}},
  \bibinfo{author}{\bibfnamefont{T.}~\bibnamefont{Ky\^omen}},
  \bibinfo{author}{\bibfnamefont{M.}~\bibnamefont{Itoh}}, \bibnamefont{and}
  \bibinfo{author}{\bibfnamefont{M.}~\bibnamefont{Oguni}},
  \bibinfo{journal}{Phys. Rev. B} \textbf{\bibinfo{volume}{69}},
  \bibinfo{pages}{144406} (\bibinfo{year}{2004}).

\bibitem[{\citenamefont{He et~al.}(2009)\citenamefont{He, Zheng, Mitchell, Foo,
  Cava, and Leighton}}]{he.apl09}
\bibinfo{author}{\bibfnamefont{C.}~\bibnamefont{He}},
  \bibinfo{author}{\bibfnamefont{H.}~\bibnamefont{Zheng}},
  \bibinfo{author}{\bibfnamefont{J.~F.} \bibnamefont{Mitchell}},
  \bibinfo{author}{\bibfnamefont{M.~L.} \bibnamefont{Foo}},
  \bibinfo{author}{\bibfnamefont{R.~J.} \bibnamefont{Cava}}, \bibnamefont{and}
  \bibinfo{author}{\bibfnamefont{C.}~\bibnamefont{Leighton}},
  \bibinfo{journal}{Appl. Phys. Lett.} \textbf{\bibinfo{volume}{94}},
  \bibinfo{pages}{102514} (\bibinfo{year}{2009}).

\bibitem[{\citenamefont{Yamaguchi et~al.}(1996)\citenamefont{Yamaguchi,
  Okimoto, Taniguchi, and Tokura}}]{yamaguchi.prb96}
\bibinfo{author}{\bibfnamefont{S.}~\bibnamefont{Yamaguchi}},
  \bibinfo{author}{\bibfnamefont{Y.}~\bibnamefont{Okimoto}},
  \bibinfo{author}{\bibfnamefont{H.}~\bibnamefont{Taniguchi}},
  \bibnamefont{and} \bibinfo{author}{\bibfnamefont{Y.}~\bibnamefont{Tokura}},
  \bibinfo{journal}{Phys. Rev. B} \textbf{\bibinfo{volume}{53}},
  \bibinfo{pages}{R2926} (\bibinfo{year}{1996}).

\bibitem[{\citenamefont{Babushkina et~al.}(2008)\citenamefont{Babushkina,
  Taldenkov, Inyushkin, Maignan, Khomskii, and Kugel}}]{babushkina.prb08}
\bibinfo{author}{\bibfnamefont{N.~A.} \bibnamefont{Babushkina}},
  \bibinfo{author}{\bibfnamefont{A.~N.} \bibnamefont{Taldenkov}},
  \bibinfo{author}{\bibfnamefont{A.~V.} \bibnamefont{Inyushkin}},
  \bibinfo{author}{\bibfnamefont{A.}~\bibnamefont{Maignan}},
  \bibinfo{author}{\bibfnamefont{D.~I.} \bibnamefont{Khomskii}},
  \bibnamefont{and} \bibinfo{author}{\bibfnamefont{K.~I.} \bibnamefont{Kugel}},
  \bibinfo{journal}{Phys. Rev. B} \textbf{\bibinfo{volume}{78}},
  \bibinfo{eid}{214432} (\bibinfo{year}{2008}).

\end{thebibliography}

\end{document}